\documentclass[12pt]{article}
\usepackage[margin=1in]{geometry}
\usepackage{setspace}
\newif\ifjaespacing \jaespacingfalse
\newcommand{\floatnote}[1]{\par\vspace{4pt}\begin{minipage}{\linewidth}\footnotesize\textit{Notes:} #1\end{minipage}}
\usepackage{amsmath,amssymb,amsthm,mathtools}
\usepackage{bm}
\usepackage[round]{natbib}
\usepackage{booktabs}
\usepackage{enumitem}
\usepackage{graphicx}
\usepackage[hidelinks]{hyperref}
\DeclareTextSymbolDefault{\textasteriskcentered}{OMS}
\DeclareTextSymbolDefault{\textdagger}{OMS}
\DeclareTextSymbolDefault{\textdaggerdbl}{OMS}
\DeclareTextSymbolDefault{\textsection}{OMS}
\DeclareTextSymbolDefault{\textparagraph}{OMS}

\theoremstyle{plain}
\newtheorem{proposition}{Proposition}

\newtheorem{corollary}{Corollary}
\theoremstyle{definition}
\newtheorem{assumption}{Assumption}
\newtheorem{definition}{Definition}
\theoremstyle{remark}
\newtheorem{remark}{Remark}
\newtheorem{oaremark}{Remark}[section]

\newcommand{\E}{\mathbb{E}}

\newcommand{\Prob}{\Pr}
\newcommand{\Phinorm}{\Phi}
\newcommand{\phinorm}{\varphi}
\newcommand{\indic}{\mathbf{1}}
\newcommand{\trans}{^{\prime}}
\newcommand{\given}{\,\vert\,}
\newcommand{\APE}{\mathrm{APE}}
\newcommand{\AIE}{\mathrm{AIE}}

\DeclareMathOperator*{\plim}{plim}
\newcommand{\independent}{\mathrel{\perp\!\!\!\perp}}
\newcommand{\tcf}{\theta_{\mathrm{cf}}}
\newcommand{\calG}{\mathcal{G}}
\newcommand{\zq}{z_{1-\alpha/2}}
\DeclareMathOperator{\Var}{Var}
\DeclareMathOperator{\Cov}{Cov}
\DeclareMathOperator{\avar}{avar}

\title{\large\textbf{Plausibly Exogenous Instruments in Probit Models:\\
Sensitivity Bounds for Average Partial and Interaction Effects}\thanks{\emph{Acknowledgments.}
The authors used Claude (Anthropic; several model versions,
May--November 2026) to write and test the Stata and R code, run the simulations and
applications, and draft and edit the text. The authors directed this work, reviewed and
verified all code, results and text, and take full responsibility for the content.
\emph{Data availability.} All data are public: the NELS:88 extract in Wooldridge's
textbook datasets (Catholic schooling), the Angrist--Evans replication files on the
Harvard Dataverse (doi:10.7910/DVN/4W9GW2), and the Chen--Ma--Sinclair replication
package on Zenodo (doi:10.5281/zenodo.5525846). The \texttt{apexog}
command is public for Stata (\url{https://github.com/jespernwulff/stata-apexog}) and R
(\url{https://jespernwulff.github.io/apexog}).}}
\author{Luke Taylor\thanks{Aarhus Center for Econometrics (ACE), Aarhus University,
\texttt{lntaylor@econ.au.dk}. Financial support is gratefully acknowledged from the
Independent Research Fund Denmark (DFF grant ID 4284-00054B) and the Aarhus Center for
Econometrics (ACE) funded by the Danish National Research Foundation grant number DNRF186.}
\and Jesper N.\ Wulff\thanks{Aarhus University, \texttt{jwulff@econ.au.dk}.}}
\date{This version: 17 November 2026}
\ifjaespacing\fi

\begin{document}
{\singlespacing\maketitle}

\begin{abstract}
\noindent
Sensitivity analysis for imperfect instruments usually targets a linear two-stage
least squares coefficient, yet many applications have binary outcomes and
probability-scale targets. We extend the plausibly exogenous framework to
the control-function probit, targeting average partial and interaction effects. We
derive the generated-regressor variance, union and local-to-zero intervals, and
their asymptotic coverage under stated model and calibration conditions.
A violation stated on the wrong latent scale can miscalibrate the interval by
a factor set by the endogeneity correlation. Applications to Catholic schooling, sibling sex
composition and Confucian clans compare probit and linear analyses. Stata and R implementations
accompany the analysis.


\end{abstract}

\section{Introduction}
\label{sec:intro}

Instrumental-variable methods identify causal effects when an instrument is excluded
from the outcome equation. In practice exclusion is rarely exact: an instrument may
have a small direct effect on the outcome. \citet{conley_plausibly_2012} formalize
this as \emph{plausible exogeneity}, in which the exclusion violation $\gamma$ lies
in a known set or is drawn from a known prior, and translate prior beliefs about
$\gamma$ into bounds on the structural coefficient of a linear model. The approach is implemented in the \texttt{plausexog} command
\citep{clarke_practical_2018} and sits alongside other treatments of imperfect
instruments: partial identification \citep{nevo_identification_2012}, Bayesian
uncertainty about the exclusion restriction \citep{kraay_instrumental_2012}, and
extensions of plausible exogeneity itself \citep{van_kippersluis_beyond_2018}.

Much applied IV work has a \emph{binary} outcome and seeks an effect on the
\emph{probability} scale: an average partial effect (APE) or, when a treatment is
interacted with a moderator, an average interaction effect (AIE). This paper extends
the sensitivity machinery to the control-function probit. The question is what an
economically credible departure from the exclusion restriction does to the
substantive conclusion: how large would a direct effect of the instrument have to
be before the estimated effect loses significance or changes sign?

A tempting alternative is to run \texttt{plausexog} on the binary outcome as a linear probability model (LPM). Under certain conditions, linear IV and a correctly specified control-function probit estimate the same population-average effect. In the model with one continuous endogenous regressor and a linear first stage, this equality holds under exact exclusion when the instrument is Gaussian and independent of the structural disturbances, subject to the usual regularity conditions. The linear-IV coefficient then equals the population APE, although the two estimators need not give identical estimates or standard errors in a finite sample.

With covariates appropriately controlled for ($\E[z_i\mid x_i]$ linear), the comparison also depends on
weighting. If the instrument is conditionally Gaussian and its first-stage
coefficient is constant, linear IV weights the conditional average effects by
the instrument's conditional variance. Constant conditional instrument variance
is therefore a sufficient additional condition for equality with the
population APE. These are exact conditions; an approximately Gaussian instrument
does not guarantee equality.

Outside these conditions, the two methods can estimate different objects.
Our numerical design illustrates this possibility: a skewed instrument produces
a substantial gap between the linear-IV coefficient and the probit APE.
That gap does not by itself mean that linear IV estimates its own target
incorrectly; it means that its target can differ from the average effect
of interest here.

\paragraph{Relation to \citet{conley_plausibly_2012}.} Section~8.7 of the
working-paper version \citep{conley_plausibly_2008}, henceforth CHR, which the published article does
not include, already states both inference methods for a generic M-estimator: given a
consistent variance estimator at each assumed violation, it forms the Wald interval
there and plugs it into the union and the prior-weighted intervals, and for the
local-to-zero approach it expands the first-order condition along a drifting sequence
$\gamma_n=\nu/\sqrt n$. Those constructions are the engines for our union and
local-to-zero results. Our specialization supplies (a) the variance needed here,
namely the generated-regressor variance for probability-scale functionals of a
two-step control-function estimator, with the sensitivity matrix $A$ in closed
form; (b) sufficient conditions for prior-averaged coverage at a \emph{fixed}
prior scale; and (c) an analysis of the structural-versus-index units hazard.
Propositions~\ref{prop:normal}--\ref{prop:ltz} supply (a) and specialize CHR's
generic constructions to this estimator; Proposition~\ref{prop:plateau} states
the additional fixed-prior conditions and the implications of scale miscalibration.

\paragraph{Contributions.} (i) We extend plausible exogeneity to the control-function
probit, targeting the APE and the AIE rather than a coefficient, with a two-step
variance that accounts for the estimated control function both through the fitted
coefficients and directly through the effect calculation
(Section~\ref{sec:est}). (ii) For the union of confidence intervals (UCI) and the
local-to-zero (LTZ) interval we give a coverage theory
(Section~\ref{sec:inference}): the continuum UCI has pointwise asymptotic coverage
of at least $1-\alpha$. The LTZ interval has prior-averaged asymptotic exactness
along the local sequence and, under uniformity and monotonicity conditions, in
the prior-dominated limit. (iii) We apply the methods to three published designs
with a binary outcome and a contested instrument, running the linear and the
probit analyses using explicitly stated probability-scale calibrations
(Section~\ref{sec:app}). The applications concern Catholic schooling, sibling sex
composition and Confucian clans, and distinguish sensitivity of the significance
from sensitivity of the sign. The \texttt{apexog} implementations in Stata and R
make the methods available for applied work.

Along the way we document an elicitation issue: the violation $\gamma$ can be
stated on the \emph{structural} scale of the latent outcome or on the \emph{index}
scale of the estimating probit. A prior stated on one scale and consumed on the
other is miscalibrated. We quantify the coverage implications in a Gaussian
linear experiment, and Monte Carlo experiments illustrate the
effect of restating the prior in offset units (Section~\ref{sec:mc}).
Probability-scale elicitation, which we recommend, is more directly interpretable
but also requires an explicit conversion rule.

\paragraph{Related literature.} We build on control functions for nonlinear models
\citep{rivers_limited_1988, blundell_endogeneity_2004, blundell_endogeneity_2003,
wooldridge_control_2015, terza_two_stage_2008}, on interaction effects in nonlinear
models \citep{ai_interaction_2003, greene_testing_2010, puhani_treatment_2012,
radean_ginteff_2023}, and on two-step asymptotics
\citep{pagan_econometric_1984, murphy_estimation_1985, newey_large_1994}.

\section{Model and estimands}
\label{sec:model}

\subsection{A triangular probit model with a possibly invalid instrument}

For unit $i=1,\dots,n$ let $y_i\in\{0,1\}$ be a binary outcome, $d_i$ an endogenous
regressor, $x_i$ a $k_x\times1$ vector of exogenous covariates (including a constant),
and $z_i$ a $k_z\times1$ vector of instruments. The structural model is triangular:
\begin{align}
y_i &= \indic\{\, x_i\trans\beta + \delta\, d_i + z_i\trans\gamma + \varepsilon_i > 0 \,\},
\label{eq:struct}\\
d_i &= w_i\trans\pi + v_i, \qquad w_i \equiv (x_i\trans, z_i\trans)\trans,
\label{eq:fs}
\end{align}
where $\gamma$ is the \emph{exclusion-restriction violation}: the direct effect of the
instrument on the latent index. Valid exclusion is the point hypothesis $\gamma=0$;
plausible exogeneity replaces it with prior information on $\gamma$
(Section~\ref{sec:prior}). The single endogenous regressor is for exposition;
Online Appendix~\ref{app:firststage} treats multiple endogenous regressors and the binary-$d$ and nonlinear
first-stage cases.

\begin{assumption}[Triangular control-function structure]\label{a:cf}
$(\varepsilon_i, v_i)$ is independent of $w_i$, jointly normal and centered,
with $\sigma_v^2>0$, $|\rho|<1$, and $\Var(\varepsilon_i)=1$ (the probit normalization). Write
$\rho=\mathrm{Corr}(\varepsilon_i,v_i)$ and $\sigma_v^2=\Var(v_i)$. Then
$\E[\varepsilon_i \given v_i] = (\rho/\sigma_v)\, v_i$ and
$\Var(\varepsilon_i\given v_i)=1-\rho^2 \equiv \tilde\sigma^2$.
\end{assumption}

Under Assumption~\ref{a:cf}, write
$\varepsilon_i = (\rho/\sigma_v) v_i + e_i$ with
$e_i \perp (v_i, w_i)$ and $e_i\sim N(0,\tilde\sigma^2)$. Conditioning on the control
function $v_i$ and dividing the latent index by $\tilde\sigma$,
\begin{equation}
\Prob(y_i = 1 \given x_i, d_i, z_i, v_i)
= \Phinorm\!\big( x_i\trans\beta + \delta d_i + z_i\trans\gamma + \tcf\, v_i \big),
\qquad \tcf \equiv \frac{\rho}{\sigma_v\,\tilde\sigma},
\label{eq:cfprobit}
\end{equation}
where, with a slight abuse of notation, $(\beta,\delta,\gamma)$ now denote the
structural coefficients scaled by $1/\tilde\sigma$, the scale of the probit index, and
$\tcf$ is the control-function coefficient. The data do not identify $\gamma$: because
$d_i=w_i\trans\pi+v_i$, a shift of $\gamma$ along $\pi_z$ (any shift, with one instrument)
is absorbed by $(\beta,\delta,\tcf)$ without changing the index. Fixing $\gamma$ is an
identifying restriction, not a normalization; given $\gamma$ and the rank conditions the
remaining coefficients are identified, and all effects below are functions of them and of
the first-stage parameters, invariant to a consistent change of normalization. The violation $\gamma$ is measured
in units of the probit index (``index units''), the natural scale for the offset in
\eqref{eq:cfprobit}, and the same violation is a number $1/\sqrt{1-\rho^2}$ times larger
on that scale than on the structural scale. Both scales are normalizations, and a
latent-scale belief is undefined until the researcher says which one it refers to;
Section~\ref{sec:coverage} prices the confusion.

\subsection{Control-function representation and estimation target}

Equation \eqref{eq:cfprobit} is the control-function probit of
\citet{rivers_limited_1988} and \citet{blundell_endogeneity_2004}: the first-stage
residual $v_i$ enters the index as an additional regressor, and its coefficient
$\tcf$ both tests and corrects for endogeneity. Because $v_i$ is unobserved it is
replaced by the estimated residual $\hat v_i = d_i - w_i\trans\hat\pi$; the resulting
two-step estimator and its generated-regressor variance are the subject of
Section~\ref{sec:est}. For a nonlinear first stage (binary, fractional, or count $d$)
a generalized residual can be used as a control only under additional,
model-specific restrictions \citep{terza_two_stage_2008, wooldridge_control_2015}.
Its sufficiency and the chosen conditional probit mean do not follow from
Assumption~\ref{a:cf};
see Online Appendix~\ref{app:firststage}.

Write $\eta_i(\gamma) \equiv x_i\trans\beta + \delta d_i + z_i\trans\gamma + \tcf v_i$
for the index at a given violation $\gamma$, and collect the second-stage parameters in
$\vartheta \equiv (\beta\trans, \delta, \tcf)\trans$.

\subsection{Estimands: average partial and interaction effects}
\label{sec:estimands}

The objects of interest are probability-scale effects, defined \emph{conditional on the
control function} (the average structural function integrates it out; Section~\ref{sec:mc-uci}), with population definitions below and sample analogues
used for estimation. Section~\ref{sec:est} distinguishes inference for the
sample functional from inference for the population average.

\begin{definition}[APE]\label{def:ape}
For a continuous endogenous regressor, the average partial effect at violation $\gamma$
is
\begin{equation}
\APE(\gamma) \;=\; \E\!\big[\, \phinorm(\eta_i(\gamma))\,\delta \,\big],
\label{eq:ape-cts}
\end{equation}
with $\phinorm$ the standard normal density. For a binary endogenous regressor it is the
discrete change
\begin{equation}
\APE(\gamma) \;=\; \E\!\big[\, \Phinorm(\eta_i^{(1)}(\gamma)) - \Phinorm(\eta_i^{(0)}(\gamma)) \,\big],
\label{eq:ape-bin}
\end{equation}
where $\eta_i^{(d)}(\gamma)$ sets $d_i=d$ in the index while holding the
control fixed. Interpreting this contrast causally requires a model for both counterfactual
outcomes at that control value; the recursive bivariate probit used below specifies one by
joint normality.
\end{definition}

\begin{definition}[AIE]\label{def:aie}
Suppose the index contains an interaction $\delta_{jk}(r_{ji}\,r_{ki})$ between two
regressors $r_j, r_k$ (at least one endogenous), with own coefficients $\delta_j,
\delta_k$. The average interaction effect is the average cross effect of
$\Phinorm(\eta_i)$ in $(r_j, r_k)$, with $\eta_i=\eta_i(\gamma)$: for two continuous regressors,
\begin{equation}
\AIE(\gamma) \;=\; \E\!\big[\, \phinorm(\eta_i)\big(\delta_{jk} - \eta_i\, a_i\, b_i\big)\big],
\qquad a_i = \delta_j + \delta_{jk} r_{ki},\;\; b_i = \delta_k + \delta_{jk} r_{ji};
\label{eq:aie-ctscts}
\end{equation}
for a continuous-by-binary pair it is the first difference of the continuous slope across
the binary level, and for a binary-by-binary pair the second difference of
$\Phinorm(\eta_i)$ \citep[cf.][]{ai_interaction_2003, radean_ginteff_2023}. The three
forms are given explicitly in Online Appendix~\ref{app:comp}.
\end{definition}

Write $O_i=(y_i,x_i,d_i,z_i)$ for the observed data and
$v_i(\pi)=d_i-w_i\trans\pi$. Each effect is a smooth functional
$\theta(\gamma)=\E[q(O_i,\vartheta,\pi,\gamma)]$, with the dependence on
$\pi$ entering through the control. Differentiation with respect to treatment
holds that control fixed; differentiation with respect to the estimated parameters
must account for its dependence on $\pi$.

\subsection{Prior information on the violation}
\label{sec:prior}

Plausible exogeneity replaces $\gamma=0$ with one of two forms of prior information,
mirroring \citet{conley_plausibly_2012}:
\begin{enumerate}[label=(\roman*),nosep]
\item \emph{Bounded support} (for the UCI): $\gamma\in\calG$, with $\calG$ a known bounded set,
typically a box $\prod_\ell[\underline\gamma_\ell,\overline\gamma_\ell]$.
\item \emph{A distribution} (for the LTZ): $\gamma\sim F$, e.g.\ $F=N(\mu,\Omega)$ or a
uniform, capturing a researcher's belief about the magnitude and sign of the violation.
\end{enumerate}
To first order, one index unit of $\gamma$ shifts $\Prob(y=1)$ by
$\bar m \equiv \E[\phinorm(\eta_i)]$ probability points per unit of $z$, so a
probability-point prior can be approximately rescaled by $1/\bar m$: the
\texttt{gunits(pr)} option of \texttt{apexog} uses the baseline fitted value
(Online Appendix~\ref{app:alg}). This is a local conversion; for a binary instrument the
exact direct effect is a finite probability difference (Section~\ref{sec:app}).

\section{Estimation and two-step asymptotics}
\label{sec:est}

\subsection{The two-step control-function estimator}

For a \emph{known} violation $\gamma$, estimation proceeds in two steps.

\emph{First stage.} $\hat\pi$ is the OLS estimator in \eqref{eq:fs}, solving the sample
analog of $\E[w_i(d_i - w_i\trans\pi)]=0$. Its influence function is
$\psi_{1i} = M^{-1} w_i v_i$ with $M \equiv \E[w_i w_i\trans]$, so
$\sqrt n(\hat\pi-\pi) = n^{-1/2}\sum_i \psi_{1i} + o_p(1)$ and
$V_1 \equiv \avar\{\sqrt n(\hat\pi-\pi)\} = M^{-1}\E[w_iw_i\trans v_i^2]M^{-1}$.

\emph{Second stage.} Given $\hat\pi$, form $\hat v_i = d_i - w_i\trans\hat\pi$ and the
index $\eta_i = x_i\trans\beta + \delta d_i + z_i\trans\gamma + \tcf\hat v_i$, and let
$g_i = (x_i\trans, d_i, \hat v_i)\trans$ be the second-stage regressor vector. The
probit (quasi-)score is
\begin{equation}
s_i(\vartheta,\pi) \;=\; \lambda(\eta_i)\,\big(y_i - \Phinorm(\eta_i)\big)\, g_i,
\qquad \lambda(\eta) \equiv \frac{\phinorm(\eta)}{\Phinorm(\eta)\,(1-\Phinorm(\eta))},
\label{eq:score}
\end{equation}
and $\hat\vartheta(\gamma)$ solves $n^{-1}\sum_i s_i(\hat\vartheta,\hat\pi)=0$. The offset
$z_i\trans\gamma$ enters \eqref{eq:score} through $\eta_i$ only; thus the entire family
$\{\hat\vartheta(\gamma)\}_\gamma$ is obtained by refitting the probit with the offset
shifted, holding the first stage fixed.

The estimator $\hat\vartheta(\gamma)$ is an M-estimator with a \emph{generated regressor}
$\hat v_i$, whose presence, through its dependence on $\hat\pi$, alters the asymptotic
variance relative to a probit that treats $\hat v_i$ as data
\citep{pagan_econometric_1984}.

\subsection{Asymptotic normality with a Murphy--Topel variance}

The regularity conditions R1--R5 (Online Appendix~\ref{app:cond}) are i.i.d.\
sampling with finite moments, Assumption~\ref{a:cf}, identification (including
instrument relevance, $\pi_z\ne0$, and a nonsingular augmented probit information),
smoothness with dominated moments, and interior, consistent two-step estimation.
Define the (expected) second-stage Hessian and the cross-derivative of the
second-stage score in the first-stage parameter,
\begin{equation}
H \;\equiv\; \E\!\left[\frac{\partial s_i(\vartheta,\pi)}{\partial\vartheta\trans}\right]
= -\,\E\!\big[\omega(\eta_i)\, g_i g_i\trans\big],
\qquad
G \;\equiv\; \E\!\left[\frac{\partial s_i(\vartheta,\pi)}{\partial\pi\trans}\right],
\label{eq:HG}
\end{equation}
with the probit weight $\omega(\eta)\equiv \phinorm(\eta)^2/[\Phinorm(\eta)(1-\Phinorm(\eta))]$.
The matrix $G$ collects the two channels through which $\hat\pi$ enters
$s_i$, the index and the $\hat v_i$ entry of $g_i$; its closed form is in Online
Appendix~\ref{app:cond}.

\begin{proposition}[Two-step normality]\label{prop:normal}
Under R1--R5 and a fixed, known true violation $\gamma$,
\[
\sqrt n\,\big(\hat\vartheta(\gamma)-\vartheta(\gamma)\big)\;\xrightarrow{\;d\;}\;N\!\big(0,\,V_\vartheta\big),
\qquad
V_\vartheta = H^{-1}\,\Sigma\,H^{-\prime},\quad
\Sigma \equiv \E\!\big[(s_i + G\psi_{1i})(s_i + G\psi_{1i})\trans\big].
\]
Equivalently, with $S=\E[s_is_i\trans]$ (the naive probit meat),
$C=\E[s_i\psi_{1i}\trans]$, and $V_1=\E[\psi_{1i}\psi_{1i}\trans]$,
\[
V_\vartheta = H^{-1}\big(S + G V_1 G\trans + G C\trans + C G\trans\big)H^{-\prime},
\]
the two-step \citet{murphy_estimation_1985} variance. The naive probit variance
$H^{-1} S H^{-\prime}$ omits the $G\psi_{1}$ terms and is, in general, inconsistent for
$V_\vartheta$.
\end{proposition}

\subsection{Effects as smooth functionals and their delta-method variance}

For fixed index-scale $\gamma$, the estimator and sample target are
\begin{align}
\hat\theta(\gamma)&=n^{-1}\sum_i
 q(O_i,\hat\vartheta(\gamma),\hat\pi,\gamma),\\
\theta_n(\gamma)&=n^{-1}\sum_i
 q(O_i,\vartheta(\gamma),\pi,\gamma).
\end{align}
The second expression evaluates the effect at the realized observations and
true parameters, including the true first-stage control; we retain this target for
the main coverage statements and Monte Carlo comparisons. The population target
$\theta(\gamma)=\E[q(O_i,\vartheta(\gamma),\pi,\gamma)]$ additionally includes
uncertainty from averaging over the population.

Let $\psi_{2i}=-H^{-1}(s_i+G\psi_{1i})$ be the second-stage influence function
from Proposition~\ref{prop:normal}, and define column gradients
\begin{equation}
 J_\vartheta=\E[\partial_\vartheta q],\qquad
 J_\pi=\E[\partial_\pi q].
\label{eq:jac}
\end{equation}
The derivative $J_\pi$ holds $\vartheta$ and $\gamma$ fixed. The first stage
therefore enters both through $\psi_{2i}$ and directly through $J_\pi$.

\begin{proposition}[Full delta-method variance of an effect]\label{prop:delta}
Under R1--R5 at a fixed true $\gamma$, with joint differentiability, an integrable envelope for the derivatives of
$q$ in $(\vartheta,\pi)$, finite second moments of the stated influence
functions and positive limiting variance,
\begin{align}
\sqrt n\{\hat\theta(\gamma)-\theta_n(\gamma)\}
 &=n^{-1/2}\sum_i\ell_{Si}+o_p(1),&
\ell_{Si}&=J_\vartheta\trans\psi_{2i}+J_\pi\trans\psi_{1i},\label{eq:if-sample}\\
V_\theta\equiv V_S&=\E[\ell_{Si}^2]
 =J_\vartheta\trans V_\vartheta J_\vartheta
  +J_\pi\trans V_1J_\pi
  +2J_\vartheta\trans V_{21}J_\pi,&
V_{21}&=-H^{-1}(C+GV_1).\label{eq:full-effect-var}
\end{align}
For the population target the influence function and variance are instead
\begin{equation}
 \ell_{Pi}=q(O_i,\vartheta,\pi,\gamma)-\theta(\gamma)+\ell_{Si},
 \qquad V_P=\E[\ell_{Pi}^2].\label{eq:if-population}
\end{equation}
For either target the centered error is asymptotically normal with the
corresponding variance. Consistent variance estimates are the empirical
variances of the plug-in influence functions; the squared standard error is
$\hat V_S/n$ or $\hat V_P/n$, respectively.
\end{proposition}

The proof is a joint Taylor expansion in $(\vartheta,\pi)$ (Online Appendix~\ref{app:proofs13}).
Note that $V_P$ is \emph{not} generally $V_S+\Var(q_i)$: it also contains
$2\Cov(q_i,\ell_{Si})$. Nor is a Murphy--Topel correction to $V_\vartheta$
alone sufficient for either target when $J_\pi\ne0$; the extra terms can
increase or decrease the variance. The standard error of \texttt{margins} after
\texttt{cfprobit} omits the direct term as well; \texttt{apexog} includes it. At
correctly specified offsets and in the applications the term changes the standard error
by about $1\%$ and at most by about $3\%$ (Online Appendix~\ref{app:agree}); at a
misspecified offset it reaches $9\%$ in one Poisson cell (Section~\ref{sec:mc}).

For the continuous APE, $q_i=\delta\phinorm(\eta_i)$, the required gradients are
\begin{equation}
 J_\vartheta=\E[\delta\phinorm'(\eta_i)g_i]
              +\E[\phinorm(\eta_i)]e_\delta,
 \qquad J_\pi=-\delta\tcf\,\E[\phinorm'(\eta_i)w_i],
 \label{eq:ape-full-jac}
\end{equation}
where $e_\delta$ selects the treatment coefficient. For the binary contrast
holding a common control fixed in both arms,
$J_\pi=-\tcf\E[(\phinorm(\eta_i^{(1)})-
\phinorm(\eta_i^{(0)}))w_i]$. Online Appendix~\ref{app:jac} gives the direct gradient for every target and for vector or
nonlinear controls.

Under clustered sampling the influence functions are summed within clusters (Online
Appendix~\ref{app:jac}). The recursive bivariate probit is a joint maximum-likelihood estimator, so its influence function has no
two-step terms.

\paragraph{Counterfactual offsets.}
Propositions~\ref{prop:normal}--\ref{prop:delta} are stated at the true violation. At a
different offset, regular misspecified-M-estimation theory yields the same stacked-influence
construction about a pseudo-true target, using the full score derivatives rather than the
information identity, as \texttt{apexog} does (Online Appendix~\ref{app:jac}). Such a sandwich
estimates uncertainty
about that offset's pseudo-true effect, not its bias relative to the true causal effect. At the
true offset the reported standard error matches the sampling standard deviation in every two-step
design we simulate (ratios $0.96$--$1.03$). At the $\gamma_0=0$ fit of the linear design with the truth
$0.2$ or $0.5$ structural units away, the sampling standard deviation exceeds it by $5\%$ and
$22\%$ but the population-target standard error of \eqref{eq:if-population} by under $1\%$:
away from the true offset the two targets separate (Section~\ref{sec:mc-uci}).

\section{Inference under plausible exogeneity}
\label{sec:inference}

Fix a target effect $\theta$ (an APE or an AIE). Refitting the control-function probit
with the offset $z_i\trans\gamma_0$ and applying
Propositions~\ref{prop:normal}--\ref{prop:delta} delivers, for every \emph{assumed}
violation $\gamma_0$, an estimate $\hat\theta(\gamma_0)$ and a standard error
$\hat\sigma(\gamma_0)$. This section turns that family into inference on the
\emph{true} effect under the two forms of prior information of
Section~\ref{sec:prior}. Throughout, $\gamma$ denotes the true violation and
\begin{equation}
m(\gamma_0;\gamma) \;\equiv\; \plim_{n\to\infty}\,\hat\theta(\gamma_0)
\quad\text{when the data are generated under violation } \gamma
\label{eq:map}
\end{equation}
is the \emph{violation-to-estimate map}: the pseudo-true value of the plug-in effect
when the researcher assumes $\gamma_0$ but the truth is $\gamma$
\citep[cf.\ the misspecified-moment analysis of][]{newey_generalized_1985}. Assuming the
right violation recovers the true effect, $m(\gamma;\gamma)=\theta(\gamma)$, and the map
is continuously differentiable near the origin under R4. We write
\begin{equation}
A \;\equiv\; -\,\frac{\partial m(\gamma_0;\gamma)}{\partial\gamma_0}\bigg|_{(0,0)}
\label{eq:A}
\end{equation}
for the (sign-flipped) sensitivity of the estimate to the assumed offset, the nonlinear
analog of the $A$ matrix of \citet{conley_plausibly_2012}; Online
Appendix~\ref{app:A} gives it in closed form. The first-stage fit does not change with the
assumed offset, so its estimation uncertainty enters $V_\theta$, not $A$.

\subsection{Union of confidence intervals (UCI)}
\label{sec:uci}

Under bounded-support prior information $\gamma\in\calG$, define for each $\gamma_0\in\calG$
the $\gamma_0$-specific Wald interval
$\mathrm{CI}_n(\gamma_0) = [\,\hat\theta(\gamma_0) \pm z_{1-\alpha/2}\,\hat\sigma(\gamma_0)\,]$
and the union interval
\begin{equation}
\mathrm{UCI}_n \;=\; \Big[\; \min_{\gamma_0\in\calG}\big\{\hat\theta(\gamma_0) - z_{1-\alpha/2}\hat\sigma(\gamma_0)\big\},\;
\max_{\gamma_0\in\calG}\big\{\hat\theta(\gamma_0) + z_{1-\alpha/2}\hat\sigma(\gamma_0)\big\} \;\Big],
\label{eq:uci}
\end{equation}
the interval hull of $\bigcup_{\gamma_0\in\calG}\mathrm{CI}_n(\gamma_0)$, which is what
\texttt{apexog} reports.

\begin{proposition}[UCI coverage]\label{prop:uci}
Under R1--R5, if the true violation satisfies $\gamma\in\calG$, then, for the sample
functional $\theta_n$ of Proposition~\ref{prop:delta},
\[
\liminf_{n\to\infty}\;\Prob\big\{\, \theta_n(\gamma) \in \mathrm{UCI}_n \,\big\} \;\ge\; 1-\alpha .
\]
\end{proposition}

The guarantee is pointwise in each fixed true violation and asymptotic: the continuum union
contains the valid Wald interval at the true violation. It is conditional on the maintained
model and regular identification, a uniform-in-$\gamma$ guarantee requires uniform regularity,
and because the software approximates the extrema on a finite grid the grid must be fine enough
for those extrema to be accurate at the $\sqrt n$ scale (Online Appendix~\ref{app:proofs13}).

\subsection{Local-to-zero (LTZ) inference}
\label{sec:ltz}

The sharper alternative treats the violation as a draw from a prior $F$ and follows
the local-to-zero device of \citet[\S 4.3 and \S 8.7]{conley_plausibly_2008}: the
violation is modeled as being of the same order as sampling uncertainty,
\[
\text{(R6)}\qquad \gamma_n = \nu/\sqrt n, \qquad \nu\sim F,\quad \E\|\nu\|^2<\infty,
\]
with $\nu$ drawn independently of the primitive covariates and disturbances
before generating the outcomes under $\gamma_n$. The outcomes depend on $\nu$
through $\gamma_n$. R6 also requires an influence expansion and a conditional
central limit theorem, uniform on each bounded set of $\nu$, with the same
limiting $N(0,V_\theta)$ law. The limiting normal noise is then independent of $\nu$.
Instrument strength remains fixed; this is not a weak-instrument result.

\begin{proposition}[Local-to-zero convolution law]\label{prop:ltz}
Under R1--R6, the estimator computed at the assumed violation $\gamma_0=0$ satisfies
\[
\sqrt n\,\big(\hat\theta(0) - \theta_n(\gamma_n)\big) \;\xrightarrow{\;d\;}\;
N(0, V_\theta) \;+\; A\,\nu,
\]
with the two terms independent, $V_\theta$ from Proposition~\ref{prop:delta} and $A$
from \eqref{eq:A}. Consequently the large-sample law of $\hat\theta(0)$ around the true
effect is the convolution $N(0,V_\theta/n) * (A\cdot F/\sqrt n)$.
\end{proposition}

The proof (Online Appendix~\ref{app:proof-ltz}) linearizes the second-stage
first-order condition along the drifting sequence, combining the nonlinear extension
in \citet[\S 8.7]{conley_plausibly_2008} with the two-step expansion behind
Proposition~\ref{prop:normal}. Conditioning on $\nu$ gives the common normal
limit on bounded sets; truncating the tails of $F$ then gives joint convergence
with $\nu$ and the independence asserted in Proposition~\ref{prop:ltz}.

\emph{Implementation.} \texttt{apexog} does not linearize: it draws
$\gamma^{(j)}\sim F$, $j=1,\dots,B$, refits at each draw to obtain
$(\hat\theta(\gamma^{(j)}), \hat\sigma(\gamma^{(j)}))$, convolves with sampling noise,
$\theta^{*(j)} = \hat\theta(\gamma^{(j)}) + \hat\sigma(\gamma^{(j)})\,\epsilon_j$,
$\epsilon_j\sim N(0,1)$, and reports the $\alpha/2$ and $1-\alpha/2$ percentiles of
$\{\theta^{*(j)}\}$. To first order the refit adjusts linearly,
$\hat\theta(\gamma^{(j)})\approx\hat\theta(0)-A\gamma^{(j)}$, so the percentile
interval inverts the convolution law of Proposition~\ref{prop:ltz} and, because
each refit subtracts the assumed violation, it is the CHR pivot interval for any
prior shape; for a Gaussian prior in the linear Gaussian experiment the construction
agrees with the analytic $V+A\Omega A\trans$ interval of \citet{conley_plausibly_2012} (width ratio $0.995$
in our checks, Online Appendix~\ref{app:agree}).

\begin{corollary}[Local exactness]\label{cor:exact}
Under R1--R6 with correctly specified $F$, consistently estimated sensitivity
and variance, continuous limiting endpoint quantiles, and the number of prior
draws tending to infinity, the LTZ percentile interval has asymptotic coverage
$1-\alpha$ for $\theta_n(\gamma_n)$, averaged over both $\nu\sim F$ and sampling.
Here the refit draws use the local distribution of $\nu/\sqrt n$. This is not
a $1-\alpha$ coverage guarantee at every fixed realization of $\nu$.
\end{corollary}

This exactness (proved in Online Appendix~\ref{app:proof-ltz}) holds \emph{along the local sequence}, where the violation shrinks with
$n$. A researcher's prior does not narrow as the sample grows, so the practically
relevant question is coverage at a \emph{fixed} prior scale.

\subsection{Coverage at fixed prior scale: the units of the violation}
\label{sec:coverage}

Fix the prior $F$ (mean zero, scale $\sigma$) and let the true violation be a draw
$\gamma\sim F$, \emph{not} shrinking with $n$. In the canonical design of
Section~\ref{sec:mc} the refit profile $\gamma_0\mapsto m(\gamma_0;\gamma)$ is exactly
linear by the identity \eqref{eq:affine-profile}, and the offset sensitivity is an order of magnitude larger
than the sensitivity of the estimand itself: $|\partial_1 m|=0.233$ against
$|\theta'(0)|=0.024$ (Online Appendix~\ref{app:map}). Under the additional uniformity and monotonicity conditions below,
a correctly specified prior gives prior-averaged exactness in the noise-free
limit. Exactness at intermediate scales is a property of the Gaussian linear
experiment, not a general finite-sample result for fitted nonlinear models.
Calibration is a further source of undercoverage, easy to miss when two natural
normalizations of the latent scale coexist. Under Assumption~\ref{a:cf} the structural index
$x\trans\beta+\delta d+z\trans\gamma_s+\varepsilon$, $\Var(\varepsilon)=1$, becomes,
after conditioning on the control function and rescaling, the probit index with
violation coefficient $\gamma_s/\tilde\sigma$, where $\tilde\sigma=\sqrt{1-\rho^2}$.
The offset in \eqref{eq:cfprobit} lives on the probit-index scale, so a researcher who
elicits ``the direct effect of $z$ on the latent outcome'' in \emph{structural} units
and supplies those numbers as offsets has supplied a prior that is too narrow by
exactly $\tilde\sigma$: a \emph{scale-miscalibrated} prior with calibration ratio
$c=\tilde\sigma<1$. The next proposition prices this miscalibration, and any other.

Below, $Q^{F}_p$ denotes the $p$-quantile of a distribution $F$; $F_1$ is the
standardized prior, the law of $\gamma/\sigma_{\text{true}}$; $F_{\text{stated}}$ is
$F$ rescaled by $c$; the violation is scalar (Online Appendix~\ref{app:proof-plateau}
discusses the vector case); the number of prior draws grows with $n$
(Section~\ref{sec:mc-design} discusses finite-draw error). Part (b), like
Proposition~\ref{prop:ltz}, concerns $\theta_n(\gamma)$; parts (a) and (c) its limit $\theta(\gamma)$.

\begin{proposition}[Coverage at fixed prior scale]\label{prop:plateau}
Let R1--R5 hold and let the true violation be $\gamma\sim F$ at fixed scale
$\sigma_{\text{true}}$, drawn before the sample, whose conditional law is
the model indexed by $\gamma$. Let $F$ be continuous with unique quantiles
at the endpoints of interest. On a compact interval containing the true and
stated supports, assume uniform consistency of the refit curve for
$m(\cdot;\gamma)$, continuity of this limit, uniformly vanishing standard
errors, and $\theta_n(\gamma)\to_p m(\gamma;\gamma)$, conditional on each
true $\gamma$. Assume numerical refit errors vanish and the number of prior
draws tends to infinity. For unbounded priors these conditions must hold on
expanding compact sets with controlled tails; a fixed truncation gives only
a truncated-prior result. Suppose the researcher's stated prior is
$F_{\text{stated}}$, $F$ rescaled by $c>0$ (so $c=1$ is correct calibration), and
let the LTZ interval be built from refits at draws from the stated prior, convolved
with sampling noise.
\begin{enumerate}[label=(\alph*),nosep]
\item \textbf{(Noise-free exactness under correct calibration.)} If $c=1$ and
$\gamma_0\mapsto m(\gamma_0;\gamma)$ is continuous and strictly monotone on the
interval containing both supports for each true $\gamma$, then as $n$ and the number of draws grow
the coverage of the LTZ percentile interval converges to exactly $1-\alpha$, for
\emph{any} such map, linear or not, separable or not, symmetric $F$ or not.
\item \textbf{(Closed form under miscalibration: linear map, Gaussian.)} Suppose the
map is linear in $\gamma_0$ with slope $-A$, $F$ is Gaussian with mean zero, the sampling
noise is Gaussian, and the reported standard error is uniformly consistent over the stated
prior's support (an assumption at counterfactual offsets; see the remark following
Proposition~\ref{prop:delta}). Let the stated prior scale drift with the sample,
$\sigma_{\text{stated}}=c\,\varsigma_\nu/\sqrt n$, $\varsigma_\nu^2=\Var(\nu)$, so that this is the local
sequence R6 with $\nu\sim F$ but stated prior $cF$; equivalently, hold $\varrho$
below fixed as $n\to\infty$. Then, in the Gaussian limit experiment implied by
Proposition~\ref{prop:normal}, coverage converges to
\begin{equation}
\kappa(\varrho, c) \;=\; 2\,\Phinorm\!\left( \zq\,
\frac{\sqrt{1+\varrho^2}}{\sqrt{1+\varrho^2/c^2}} \right) - 1,
\qquad
\varrho \;\equiv\; \frac{|A|\,\sigma_{\text{stated}}}{\sqrt{V_\theta/n}},
\label{eq:kappaform}
\end{equation}
where $\varrho$ is the stated-prior spread of the estimate relative to its sampling
standard error, fixed along the sequence; at finite $n$ it is read as
$|A|\sigma_{\text{stated}}/\sqrt{V_\theta/n}$, and the formula is then an
approximation that interpolates between the local sequence and the fixed-prior
limit of part (c).
$\kappa$ decreases (for $c<1$) from $1-\alpha$ at $\varrho=0$ to the plateau
\begin{equation}
\kappa_\infty(c) \;=\; 2\,\Phinorm\!\big(\zq\, c\big) - 1,
\label{eq:plateau}
\end{equation}
which is free of $n$, of $V_\theta$, of $|A|$, and of the instrument strength.
\item \textbf{(General plateau.)} Under the same monotonicity and uniformity conditions, for general $F$ symmetric about zero, in the prior-dominated
limit ($n\to\infty$ at fixed prior scale, which drives $\varrho\to\infty$), coverage
under miscalibration $c$ converges to
$\kappa_\infty(c)=\Prob\{|U|\le c\,Q^{F_1}_{1-\alpha/2}\}$, $U\sim F_1$, again free
of every design quantity; only monotonicity of the map is used.
\end{enumerate}
\end{proposition}

The proof is in Online Appendix~\ref{app:proof-plateau}: part (a) is uniform
convergence of the fitted curve plus quantile equivariance under monotone
transformations, which reduces the event ``target inside'' to ``$\gamma$ inside its
own quantiles''; parts (b)--(c) are direct calculations. All these probabilities average over the true prior and the sampling
law conditional on its draw, and part (b) is a Gaussian limit-experiment calculation,
not an exact finite-sample theorem at arbitrary prior widths.

Part (a) is also why the issue has no counterpart in the linear model, where the
structural and estimation scales coincide. In the probit, $c=\tilde\sigma$ is
the price of conflating the two latent scales: at $\rho=0.6$,
$\kappa_\infty=2\Phinorm(\zq\times0.8)-1=0.883$; at $\rho=0.8$, $0.760$; at
$\rho=0.3$, $0.938$ (Table~\ref{tab:rhotrack}). The plateau is free of the design, and
coverage at wide priors in our simulations is flat across sample sizes (Online
Appendix~\ref{app:agree}). Flatness across instrument strengths is weaker evidence:
sensitivity and standard error both scale with $1/\pi_z$, which leaves $\varrho$ nearly
unchanged at every prior width (Online Appendix~\ref{sec:collapse}).

At the true offset, the linear Gaussian model implies
\begin{equation}
 \hat{\tilde\sigma}(\gamma_0)
 =\{1+\hat\sigma_v^2\hat\tcf(\gamma_0)^2\}^{-1/2}.
 \label{eq:sigtilde}
\end{equation}
Its sample version is consistent at the correct specification, but at a nonzero true
violation the zero-offset loading need not be consistent for the true control coefficient, so
dividing an entire structural prior by a single baseline $\hat{\tilde\sigma}(0)$ is not
generally an exact fixed-prior repair.
For a structural-scale candidate $\gamma_s$, calibration instead requires solving
$\gamma_s=\gamma_0\hat{\tilde\sigma}(\gamma_0)$. With a linear, exactly identified first stage
this needs no refit, since $\hat\tcf(\gamma_0)=\hat\tcf(0)+\gamma_0/\hat\pi_z$ by the shift
behind \eqref{eq:affine-profile}. The estimated calibration carries its own uncertainty, which
the fixed-$\gamma$ variance of Proposition~\ref{prop:delta} does not cover, and
generalized-residual first stages require their own scale derivation.

This inversion also requires a support restriction. Writing
$a=\hat\tcf(0)$, $s=\hat\sigma_v$ and $b=1/\hat\pi_z$, its right-hand side is
$t/\sqrt{1+s^2(a+bt)^2}$, a bounded and generally nonmonotone function of $t$: it
increases only on the branch through $t=0$, where $1+s^2a(a+bt)>0$. A structural
candidate with $|\gamma_s|<|\hat\pi_z|/\hat\sigma_v$ has exactly one solution, on that
branch; outside that range a candidate has no solution or two solutions with
different effects (in the design of Section~\ref{sec:mc}, $\gamma_s=-0.9$ is solved by
the offsets $-1.01$ and $-4.70$). A structural prior must be supported within the
range, or within the image of a stated branch whose choice is an additional model
restriction. An unbounded Gaussian structural prior cannot be imposed by this inversion.

\begin{remark}[APE versus AIE]\label{cor:aie}
A common additive violation in a nonlinear index does not generally cancel
from an interaction effect. Even holding fitted coefficients fixed, a binary
interaction contrast can have the form
$D(a)=\Phinorm(a+b+c+d)-\Phinorm(a+b)-\Phinorm(a+c)+\Phinorm(a)$.
Its derivative in the common shift $a$ is the corresponding four-density
contrast, generally nonzero. Refitting adds parameter changes as well.
Low AIE sensitivity in a particular design is an empirical feature of its
refit profile. A violation on an interacted instrument may affect that profile
differently, but neither case has automatic immunity to miscalibration.
\end{remark}

\subsection{Recommendation}
\label{sec:recommend}

First, state the violation's scale, instrument increment and averaging
convention explicitly. Probability-scale elicitation is the most interpretable, but its
first-order input conversion differs from the exact refit-specific value that
\texttt{apexog} reports, and an exact support or prior on that scale would have to be imposed
refit by refit.
Second, distinguish structural and conditional-index units and do not treat
baseline latent rescaling as an exact fixed-prior correction.
Third, use the UCI for pointwise asymptotic coverage under a stated support
and the maintained model, with numerical extrema controlled. Interpret LTZ
coverage as prior-averaged, under the local or fixed-prior conditions stated
above. Interaction effects require their own sensitivity analysis.

\section{Monte Carlo evidence}
\label{sec:mc}

The simulations check the sizing of the standard errors, the coverage of the UCI, and LTZ
coverage at fixed prior scale against Proposition~\ref{prop:plateau}. They use the full
variance \eqref{eq:full-effect-var}. Against the same runs with its first term only (same data and
seeds), no coverage entry moves by more than two Monte Carlo standard errors or by more than
$0.010$: the direct term adds about $1\%$ to the standard error in these designs, and $9\%$ in
one misspecified-offset Poisson cell.

\subsection{Design}
\label{sec:mc-design}

The baseline design is the triangular model
\eqref{eq:struct}--\eqref{eq:fs} with a scalar instrument:
$x_i,z_i,v_i\sim N(0,1)$ mutually independent,
$d_i = \pi_z z_i + 0.4\,x_i + v_i$,
$y_i = \indic\{-0.2 + 0.6\,d_i + 0.5\,x_i + \gamma z_i + \varepsilon_i > 0\}$ with
$\varepsilon_i = 0.6\,v_i + e_i$, $e_i\sim N(0,1-0.36)$
(endogeneity correlation $0.6$), $\pi_z=0.8$ unless varied, and $n=2{,}000$ per
replication ($n=5{,}000$ for the UCI runs). The Monte Carlo truth in each replication
with a linear first stage is the conditional-on-CF APE of Definition~\ref{def:ape}
evaluated at the realized sample and true parameters (Online
Appendix~\ref{app:estimand}); Online Appendix~\ref{app:firststage} explains the
truths used for nonlinear first stages. Coverage entries use $R=1{,}000$ replications
unless noted, so the binomial Monte Carlo standard error is about $0.010$ near the
reported rates. The LTZ intervals of Table~\ref{tab:ltz} and Figure~\ref{fig:sweep}
use $B=1{,}000$ prior draws per fit, the software default. In the Gaussian linear limit experiment the target and the noise-convolved draws
are exchangeable under correct calibration, so a percentile interval built from order statistics
$[X_{(k_1)},X_{(k_2)}]$ has coverage exactly $(k_2-k_1)/(B+1)$: $(975-25)/1{,}001=0.949$ at
$B=1{,}000$ and $(146-4)/151=0.940$ at $B=150$. With estimated sensitivities and nonlinear refits
no such universal cap holds. The cells of Table~\ref{tab:rhotrack} at $\rho\in\{0.3,0.8\}$ and
the repair and uniform-prior experiments use $150$ draws and carry about one point of additional
simulation error (Online Appendix~\ref{app:agree} measures the effect of the draw count).

\subsection{UCI coverage and standard errors}
\label{sec:mc-uci}

Online Appendix Table~\ref{tab:uci} reports, for the assumed support
$\calG=[-0.3,0.3]$, three true violations ($\gamma=0$; $\gamma=0.2$, inside $\calG$;
$\gamma=0.5$, outside it) and five designs: the linear first stage, binary $d$ under
two data-generating processes, and the fractional-probit and Poisson first stages.
In all fifteen cells the UCI covers in every replication when $\gamma\in\calG$ and in
none when the stated support is wrong. Proposition~\ref{prop:uci} guarantees only the
first, at $1-\alpha$ and for a correctly specified model; the rest is a finding of these
designs. The
informative check is the Wald interval at the true offset, which tests the
Murphy--Topel standard error at a nonzero violation directly: where the fit is
correctly specified it covers $0.946$, $0.952$ and $0.939$ (linear first stage) and
$0.951$, $0.954$ and $0.944$ (binary $d$ under the CF-sufficient design), and the
ratio of the sampling standard deviation of the estimate to the mean reported
standard error lies between $0.96$ and $1.03$ in every cell. Away from the true
offset the sampling standard deviation can exceed the sample-target standard error
(Section~\ref{sec:est}), which does not affect the UCI
guarantee, since that rests on the Wald interval at the true offset. The linear two-step is unbiased for the
conditional-on-CF APE and biased by $13\%$ for the average-structural-function derivative (Online
Appendix~\ref{app:estimand}).

\subsection{LTZ coverage: the units experiment}
\label{sec:mc-ltz}

Our LTZ experiments state the violation $\gamma$ in \emph{structural} units and
supply a prior with the \emph{same} numerical scale to the offset, which lives in
\emph{probit-index} units. This is the elicitation error of
Section~\ref{sec:coverage} committed inside the simulator: every nominally
``correctly specified'' scenario is a scale-miscalibrated one with
$c=\tilde\sigma=\sqrt{1-0.6^2}=0.8$ exactly. Table~\ref{tab:ltz} reports coverage
for a structural-units prior comparable to sampling noise ($\gamma\sim N(0,0.05^2)$,
stated $\Omega=0.05^2$), a prior-dominated case ($\gamma\sim N(0,0.2^2)$, stated
$\Omega=0.2^2$), and a prior misstated by an additional factor of four
($\gamma\sim N(0,0.2^2)$, stated $\Omega=0.05^2$, so $c=0.8/4=0.2$).

\begin{table}[t]
\centering
\caption{LTZ coverage of the true APE}
\label{tab:ltz}
\small
\begin{tabular}{@{}llcc@{}}
\toprule
Target & Scenario & Coverage & $\kappa(\varrho,c)$ \\
\midrule
APE & structural-units prior ($c=0.8$), $\sigma=0.05$ & 0.916 & 0.910 \\
APE & structural-units prior ($c=0.8$), $\sigma=0.20$ & 0.882 & 0.886 \\
APE & structural $\sigma=0.20$, stated $0.05$ ($c=0.2$) & 0.363 & 0.384 \\
\addlinespace
AIE (endog.\ interaction) & structural-units prior, $\sigma=0.01$ & 0.958 & $\approx0.95$ \\
AIE (endog.\ interaction) & structural-units prior, $\sigma=0.05$ & 0.960 & $\approx0.95$ \\
\addlinespace
APE, two instruments & structural-units prior, $\sigma=0.05$ each & 0.909 & n/a \\
APE, two instruments & UCI benchmark, box $\pm0.2$ & 0.996 & n/a \\
\bottomrule
\end{tabular}
\floatnote{Nominal level 0.95; $R=1{,}000$ replications per cell (Monte Carlo SE $\approx0.010$), $n=2{,}000$, $B=1{,}000$ prior draws per fit. $\kappa(\varrho,c)$: the closed form \eqref{eq:kappaform} with the measured $|\partial_1m|=0.233$ and $\mathrm{SE}=0.0096$; for the AIE $\varrho\approx0$ (Remark~\ref{cor:aie}). AIE and two-instrument rows: designs in Online Appendix~\ref{app:agree}. n/a: the scalar closed form does not apply; the last row is a UCI benchmark.}
\end{table}

The closed form is accurate in the plateau and in the transition. The two $c=0.8$
APE rows sit within one Monte Carlo standard error of $\kappa(\varrho,0.8)$
($0.916$ against $0.910$ and $0.882$ against $0.886$), and the compounded
misstatement comes out at $0.363$ against a prediction of $0.384$, $1.4$ Monte
Carlo standard errors below it: the formula prices scale miscalibrations of either
origin, not just the units factor. The AIE rows are nominal or slightly above
($0.958$ and $0.960$), a design-specific finding rather than the general immunity ruled out in
Remark~\ref{cor:aie}; their prior-only
sensitivity bands are only $0.003$ and $0.014$ wide against intervals of $0.060$ and
$0.062$, so these rows test the sizing of the standard error rather than the
local-to-zero mechanism.

\begin{figure}[t]
\centering
\includegraphics[width=0.78\textwidth]{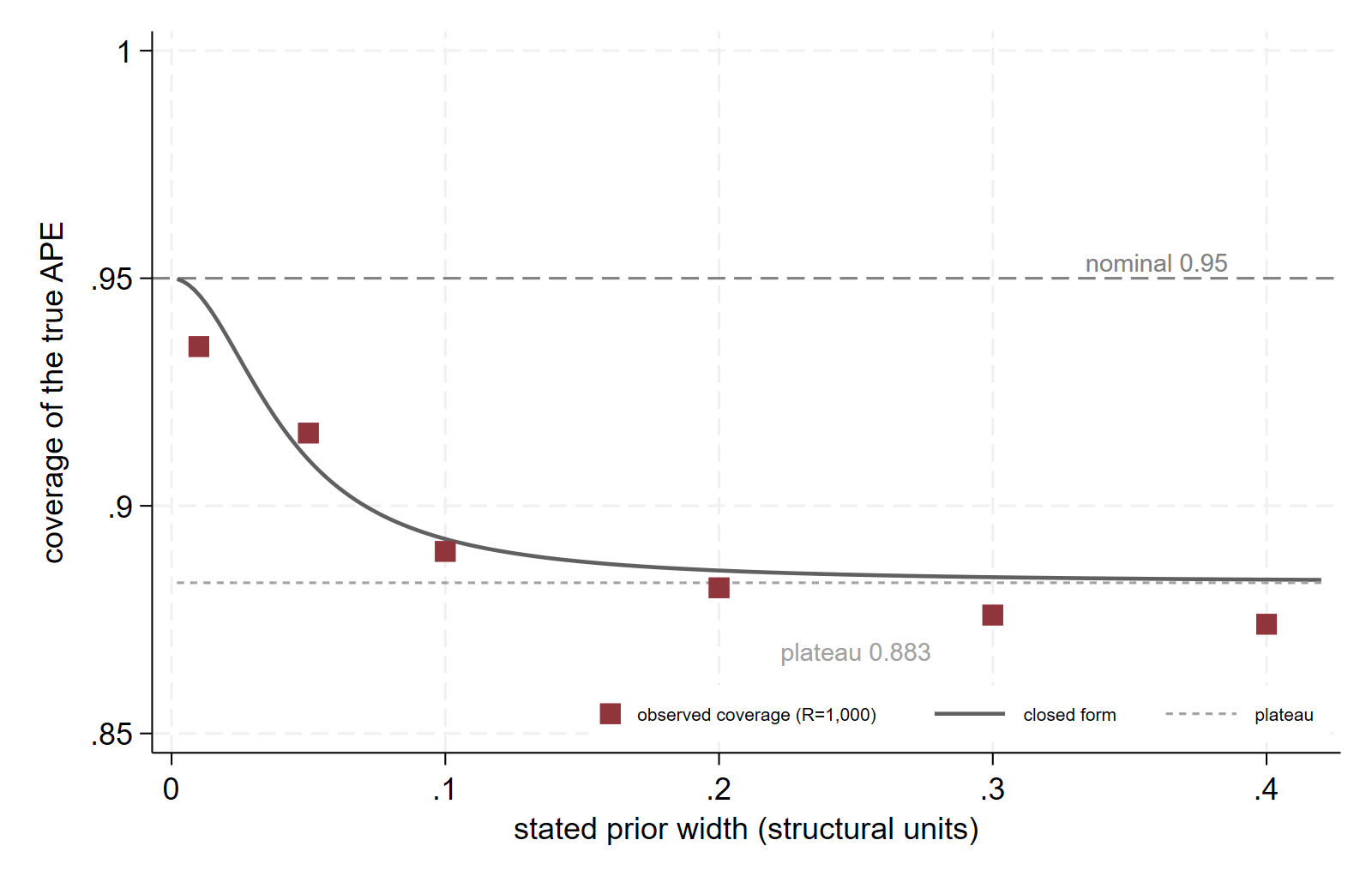}
\caption{LTZ coverage against the prior width}
\label{fig:sweep}
\floatnote{Structural-units priors, so $c=\tilde\sigma=0.8$; $R=1{,}000$ per point (Monte Carlo SE $\approx0.010$), $n=2{,}000$, $B=1{,}000$ prior draws per fit. Squares: Monte Carlo coverage; solid curve: $\kappa(\varrho(\sigma),0.8)$ from \eqref{eq:kappaform} with the inputs of Table~\ref{tab:ltz}; short dashes: the plateau $0.883$; long dashes: nominal $0.95$.}
\end{figure}

Figure~\ref{fig:sweep} traces the prior-width profile against the closed form:
coverage is near-nominal only for a near-degenerate prior ($\sigma=0.01$, where the
LTZ interval is essentially the Wald interval), then follows
$\kappa(\varrho(\sigma),0.8)$ down to its predicted plateau
$2\Phinorm(1.96\times0.8)-1=0.883$; the observed plateau over
$\sigma\in[0.1,0.4]$ averages $0.880$. The transition points sit within $0.7$ Monte
Carlo standard errors of the formula ($0.916$ against $0.910$ at $\sigma=0.05$,
$0.890$ against $0.893$ at $\sigma=0.10$); the largest residual, $0.935$ against $0.946$
at $\sigma=0.01$, is $1.5$ Monte Carlo standard errors (the Wald interval at the true
offset covers $0.943$ in the same replications).

The units mechanism carries a signature prediction: the plateau must \emph{track}
$c=\tilde\sigma=\sqrt{1-\rho^2}$ as the endogeneity correlation moves.
Table~\ref{tab:rhotrack} reruns the structural-units design at
$\rho\in\{0.3,0.6,0.8\}$, spanning plateaus from $0.938$ to $0.760$, and the plateau value
is matched at every correlation: this is the units mechanism itself, verified
against a quantity generated by the model. A \emph{uniform} structural-units prior at
$\sigma=0.30$ tests the general-$F$ plateau of Proposition~\ref{prop:plateau}(c),
whose prediction, $\Prob\{|U|\le c\,Q^{F_1}_{0.975}\}=0.8\times0.95=0.760$, differs
sharply from the Gaussian $0.883$ at the same $c=0.8$; the observed coverage is
$0.766$, so the formula captures the prior's shape, not merely its scale. The
two-instrument cell of Table~\ref{tab:ltz} ($0.909$) sits in the transition above the
scalar plateau. Its design is overidentified, so the refit map is not exactly linear and
the cell does not test the scalar reduction of Online Appendix~\ref{app:proof-plateau}.

\begin{table}[t]
\centering
\caption{The coverage plateau tracks the endogeneity correlation}
\label{tab:rhotrack}
\small
\begin{tabular}{cccccc}
\toprule
$\rho$ & $c=\tilde\sigma$ & predicted $\kappa_\infty$ & observed ($\sigma=0.15$) & observed ($\sigma=0.30$) \\
\midrule
0.3 & 0.954 & 0.938 & 0.938 & 0.930 \\
0.6 & 0.800 & 0.883 & 0.882$^{a}$ & 0.876 \\
0.8 & 0.600 & 0.760 & 0.751 & 0.770 \\
\bottomrule
\end{tabular}
\floatnote{Structural-units Gaussian priors, so $c=\tilde\sigma=\sqrt{1-\rho^2}$ and $\kappa_\infty=2\Phinorm(\zq c)-1$; nominal level 0.95, $n=2{,}000$, $R=1{,}000$ per cell (Monte Carlo SE $\approx0.010$). The $\rho=0.6$ row uses $1{,}000$ prior draws per fit, the other rows $150$ (additional quantile simulation error). $^{a}$The $\sigma=0.20$ point of the sweep in Figure~\ref{fig:sweep}, which has no $\sigma=0.15$ point.}
\end{table}

The decisive experiment is the repair: restating the prior in offset units,
$\Omega_{\text{stated}}=(\sigma/\tilde\sigma)^2$. At $\sigma=0.15$, deep in the
plateau, the oracle repair (true $\tilde\sigma=0.8$) moves coverage from $0.880$
(prediction $0.888$) to $0.952$ (prediction $0.95$; $R=250$, MC SE $0.014$). The
feasible repair, which estimates $\hat{\tilde\sigma}$ by \eqref{eq:sigtilde} in each
replication (the operation \texttt{gunits(structural)} performs), gives $0.935$,
$0.933$ and $0.942$ at $\sigma\in\{0.05, 0.15, 0.30\}$: close to nominal in aggregate but not
in each tail. The repair divides by a baseline
$\hat{\tilde\sigma}$, which is exact only at $\gamma=0$ (Section~\ref{sec:coverage}) and whose
error depends on the sign of the violation: at $\sigma=0.30$, $57$ of the $58$ misses occur
at positive violations (miss rates $0.119$ and $0.002$), and at $\sigma=0.15$, $55$ of $67$.
The rates are evidence for this design rather than a general result.

\subsection{The linear-probability shortcut}

Online Appendix~\ref{sec:mc-lpm} quantifies the cost of running \texttt{plausexog}
on the binary outcome as an LPM. With a Gaussian instrument the LPM-IV estimate and the
control-function probit both stay close to the APE even when the endogenous regressor
is heavily skewed (skewness $2.1$; errors of $-1.1\%$ and $-1.0\%$), as the conditions in the
introduction imply (Online Appendix~\ref{app:stein}). With a skewed instrument (a centered lognormal, skewness
$\approx6.2$) the LPM-IV estimate is $20\%$ below the APE at $n=200{,}000$ while the
control-function probit is within $0.4\%$, and for boxes $|\gamma|\le b$ with
$b\le0.01$ the LPM bounds exclude the true APE while the control-function probit
bounds contain it.

\section{Empirical applications}
\label{sec:app}

We apply the methods to three published instrumental-variable designs with a binary
outcome and a contested exclusion restriction, running the linear \texttt{plausexog}
analysis and the probit analysis on the same data and instrument, reporting
the assumed violation in interpretable probability units. The question we put to each design is what an economically
credible departure from the exclusion restriction does to the substantive conclusion.
For the linear two-stage least squares (2SLS) coefficient and for one or more probit
APEs we report the baseline estimate at $\gamma=0$ and two \emph{tipping values}: the
smallest direct effect of the instrument at which the $95\%$ interval first covers
zero, and the smallest at which the point estimate changes sign. The two separate a
result whose significance is fragile from one whose sign is; both are interpolated
on the union grid of Section~\ref{sec:uci}. The violation is elicited on the
probability scale, as Section~\ref{sec:recommend} recommends: $\gamma$ is the direct
effect of a unit change in the instrument on $\Prob(y=1)$, in probability points. For
the LPM that is the offset itself. For a probit we compute it exactly from the model
refitted at the violation, holding the treatment and the control function at their
observed values, as the average change in $\Prob(y=1)$ when a binary instrument
switches on, or as the average derivative for a continuous one; \texttt{apexog}
reports this value for every refit. The \texttt{gunits(pr)} option, through which the
boxes below are entered, converts at first order instead (Online
Appendix~\ref{app:apps-scale}). Box labels therefore describe the inputs to that
approximation, and the exact values achieved at refits can differ. For a binary instrument,
holding the treatment at zero rather than at its observed value defines a different
calibration, so comparisons with external benchmarks need a common convention.

The intervals below treat the baseline conversion as a plug-in calibration.
The coverage results in Section~\ref{sec:inference} take the index-scale support
or prior as specified. Inference for a fixed probability-scale violation at a nonshrinking
scale must additionally account for estimation of the conversion, through
joint influence functions or a bootstrap that repeats calibration and refitting.

For the binary-treatment applications, the CF probit rows with a linear first
stage are descriptive fitted contrasts. A binary treatment cannot satisfy
Assumption~\ref{a:cf}, and these rows carry no causal identification or causal
coverage claim. Causal interpretation of the probit-first-stage CF rows requires
the additional control-sufficiency restrictions in Online
Appendix~\ref{app:firststage}; the recursive bivariate probit instead maintains
joint normality.

\begin{figure}[tp]
\centering
\includegraphics[width=0.47\textwidth]{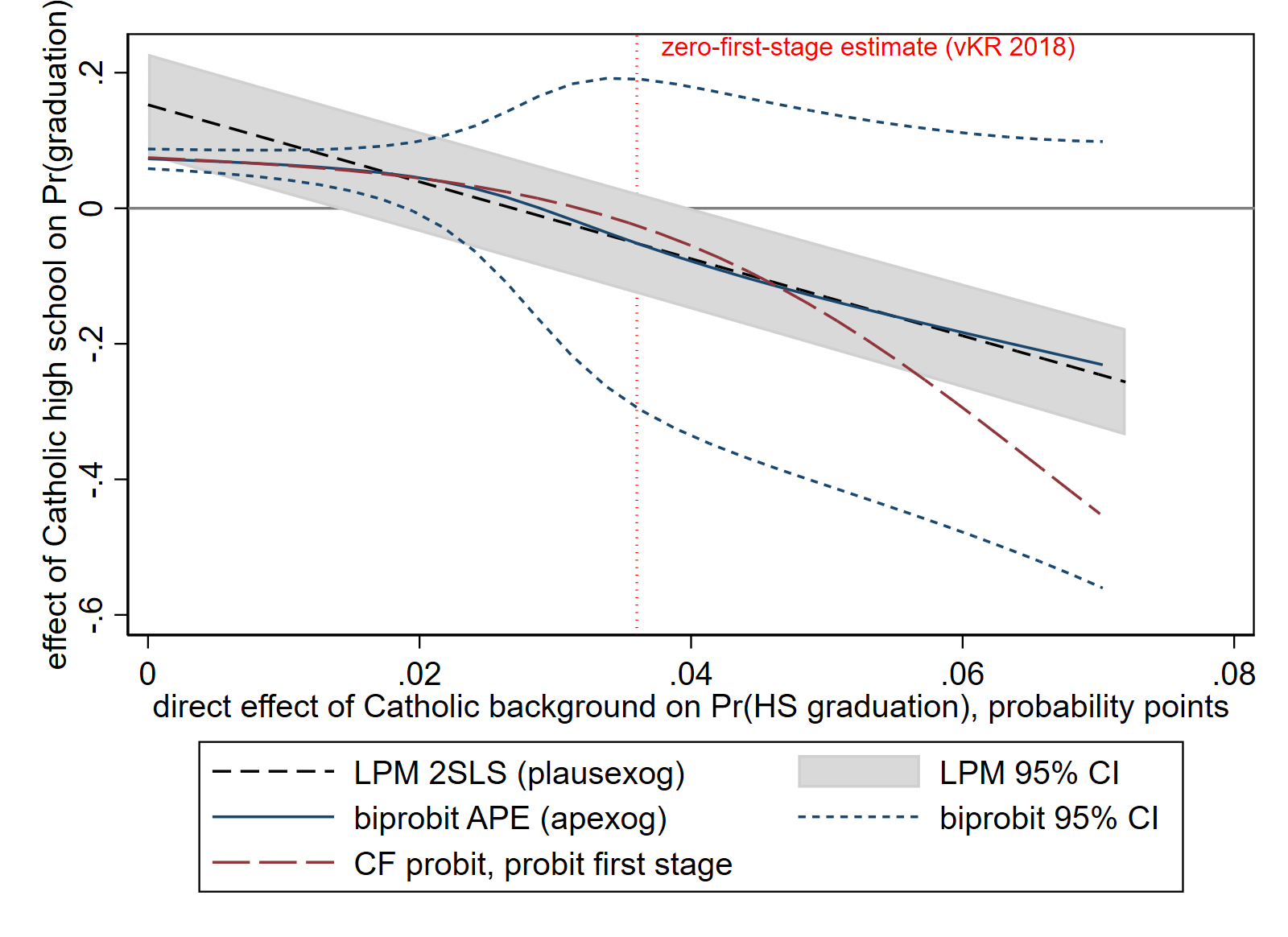}\hfill
\includegraphics[width=0.47\textwidth]{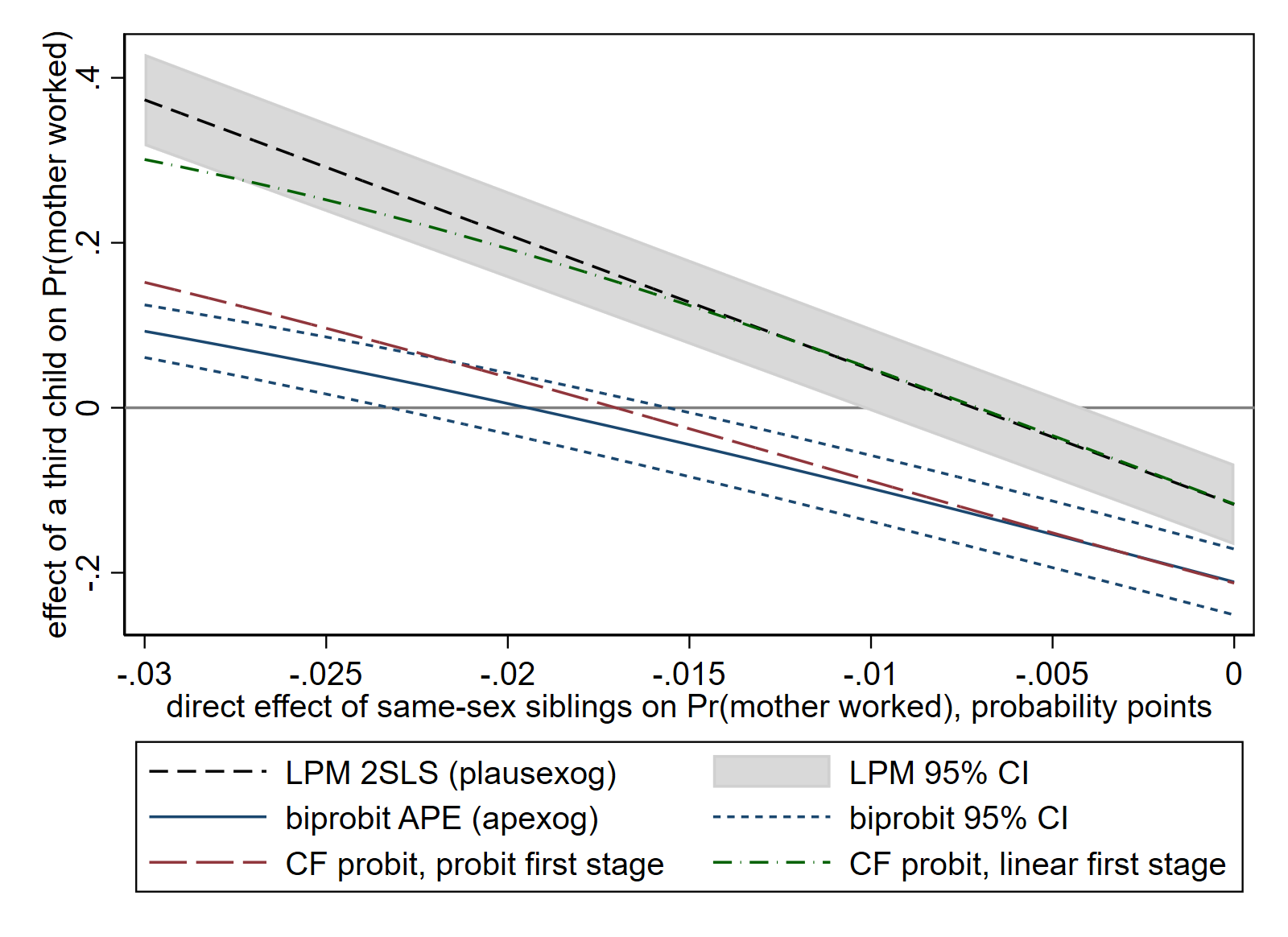}\\[4pt]
\includegraphics[width=0.47\textwidth]{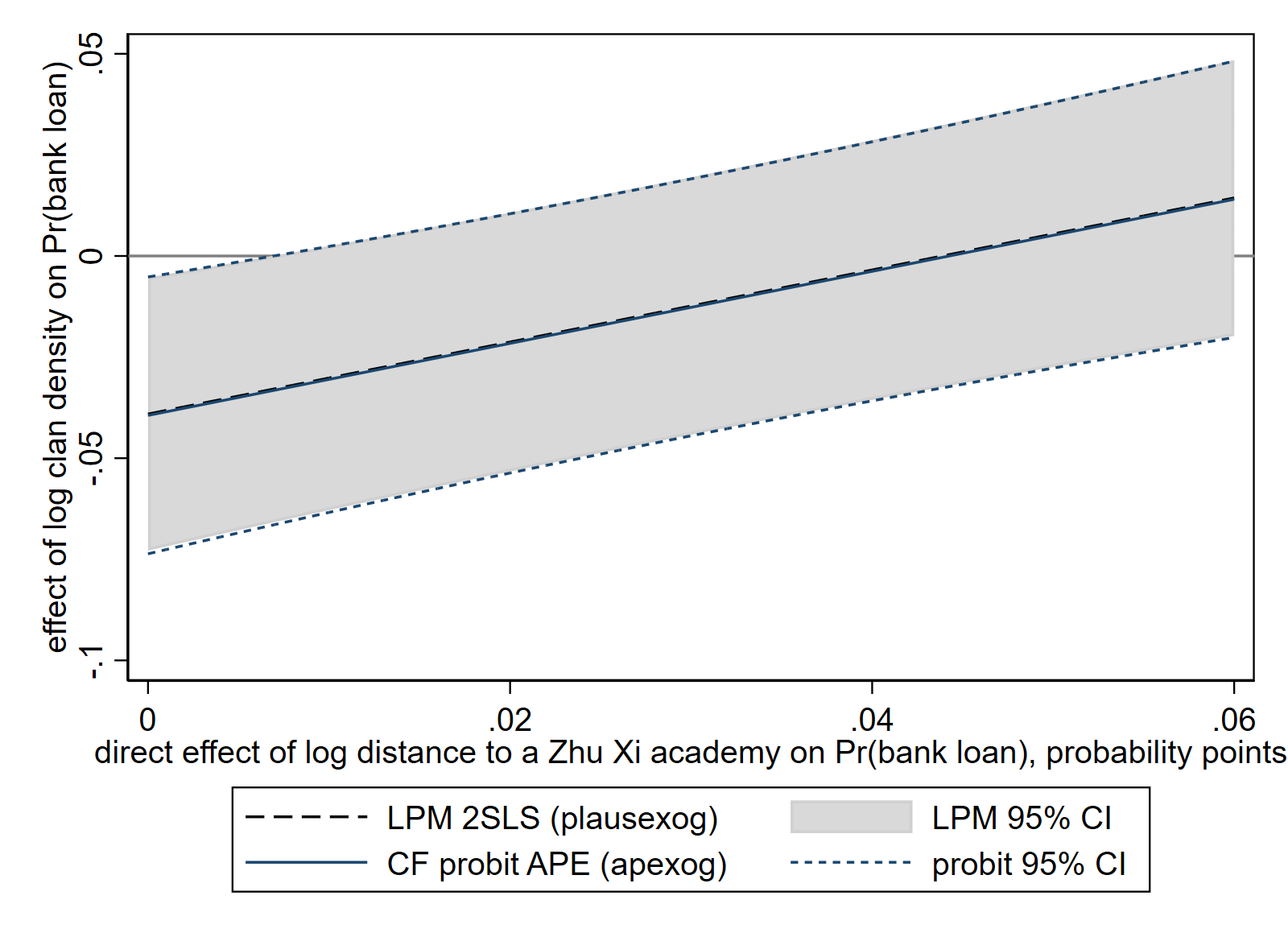}
\caption{Refit curves for the three applications}
\label{fig:apps}
\floatnote{Point estimate against the assumed direct effect $\gamma$ over the box of Table~\ref{tab:apps} (probability points, exact scale), with the $95\%$ interval of the LPM 2SLS coefficient (shaded) and of the probit APE (short dashes; the bivariate probit's in the top panels). Top left: Catholic schooling, where the shaded band reaches zero at $0.014$ and the dashed bound at $0.019$; the vertical line marks the zero-first-stage estimate of \citet{van_kippersluis_beyond_2018}. Top right: Angrist--Evans, full 1980 sample; the linear-first-stage curve lies on the LPM curve to about $-0.015$, beyond both of their tipping values. Bottom: Chen--Ma--Sinclair, violation on log distance to the nearest Zhu Xi academy.}
\end{figure}

Two facts organize the comparison. For the LPM with a single instrument the
refitted 2SLS coefficient is $\hat\beta(\gamma)=\hat\beta-\gamma/\hat\pi$, with
$\hat\pi$ the first-stage coefficient, so the sign-flip tipping value equals the
reduced-form coefficient $\hat\beta\hat\pi$ exactly, whatever the data; a linear
sensitivity analysis cannot say more about the sign than the reduced form already
does. There is also an exact identity for the continuous-treatment CF APE
with one instrument, a linear first stage and no treatment interactions:
\begin{equation}
 \hat\theta(\gamma_0)=\hat\theta(0)
 -\frac{\hat m}{\hat\pi_z}\gamma_0,\qquad
 \hat m=n^{-1}\sum_i\phinorm(\hat\eta_i(0)).
 \label{eq:affine-profile}
\end{equation}
Indeed, since $d_i=x_i\trans\hat\pi_x+z_i\hat\pi_z+\hat v_i$, shifting
$(\hat\beta,\hat\delta,\hat\tcf)$ by $(\hat\pi_x,-1,1)\gamma_0/\hat\pi_z$ absorbs the offset
and leaves all fitted indices unchanged, so the APE sign-flip value on the derivative scale is
$\hat\pi_z\hat\theta(0)$, which need not equal the LPM reduced form. The identity does not
extend to a binary-treatment discrete contrast or to the recursive bivariate probit. The second fact is the estimand
itself: with a binary treatment the probit APE is a discrete change under the
maintained model, whereas the usual complier interpretation of a binary-IV
Wald estimand requires instrument independence, exclusion, relevance and
monotonicity; the two estimands need not agree at $\gamma=0$, and when they
differ the tipping values differ with them. Table~\ref{tab:apps} collects the
results and Figure~\ref{fig:apps} draws the refit curves.

\begin{table}[tp]
\centering
\caption{Sensitivity of three published IV results to a direct effect of the instrument}
\label{tab:apps}
\footnotesize
\begin{tabular}{@{}llccc@{}}
\toprule
Application & Estimator & Baseline (SE) & CI covers 0 & Sign flips \\
\midrule
Catholic schooling & LPM 2SLS & 0.153 (0.038) & 0.014 & 0.027 \\
\quad (box $\gamma\in[0,0.072]$) & Bivariate probit & 0.073 (0.007) & 0.019 & 0.029 \\
 & CF probit, probit first stage & 0.075 (0.010) & 0.019 & 0.032 \\
 & CF probit, linear first stage$^{d}$ & 0.085 (0.010) & 0.024 & 0.034 \\
\addlinespace
Angrist--Evans & LPM 2SLS & $-0.117$ (0.025) & 0.004 & 0.007 \\
\quad (box $\gamma\in[-0.03,0]$) & Bivariate probit & $-0.211$ (0.020) & 0.016 & 0.019 \\
 & CF probit, probit first stage & $-0.213$ (0.021) & 0.014 & 0.017 \\
 & CF probit, linear first stage$^{d}$ & $-0.117$ (0.025) & 0.004 & 0.007 \\
\addlinespace
Chen--Ma--Sinclair & LPM 2SLS & $-0.039$ (0.017) & 0.007 & 0.044 \\
\quad (box $\gamma\in[0,0.06]$) & CF probit, linear first stage & $-0.039$ (0.017) & 0.007 & 0.044 \\
\bottomrule
\end{tabular}
\floatnote{$\gamma$: direct effect of a unit change in the instrument on $\Prob(y=1)$, in probability points (exact scale for the probits; Section~\ref{sec:app}); boxes entered through \texttt{gunits(pr)}. CI covers 0 (sign flips): smallest $|\gamma|$ at which the $95\%$ interval contains zero (the point estimate changes sign). Standard errors: robust for 2SLS and the bivariate probit, the full two-step variance of Proposition~\ref{prop:delta} for the control-function (CF) probit; clustered by prefecture for Chen--Ma--Sinclair. $^{d}$Descriptive fitted contrasts for binary treatments, without a causal interpretation or causal coverage claim (Section~\ref{sec:app}).}
\end{table}

\subsection{Catholic schooling and high-school graduation}
\label{sec:app-catholic}

\citet{altonji_evaluation_2005} evaluate the student's Catholic background as an
instrument for attendance at a Catholic high school, the question their companion
study \citep{altonji_selection_2005} approaches without an instrument, and
conclude that the exclusion restriction fails: a Catholic background predicts
outcomes among public-school eighth graders, almost none of whom go on to a
Catholic high school. They also ask whether a bivariate probit identifies the
effect through the exclusion restriction or through the nonlinearity of the model,
a question our bivariate-probit row inherits; \citet{han_estimation_2019} show that
identification of such a model can fail without an excluded instrument and find its
average treatment effect sensitive to the assumed joint distribution.
\citet{van_kippersluis_beyond_2018}
turn that observation into a calibrated plausibly-exogenous analysis: the
reduced-form coefficient in the zero-first-stage group estimates the direct effect,
and its standard error sets the prior variance. For high-school graduation their
direct effect is $0.036$ (standard error $0.008$) on the probability scale, the
currency Section~\ref{sec:recommend} recommends. Running \texttt{plausexog} on the
linear probability model, they find that the effect of Catholic schooling, $0.251$
in their 2SLS, falls to $0.012$ (standard error $0.046$) under that prior: the
positive effect disappears.

We use the public NELS:88 extract distributed with Wooldridge's textbook datasets
($n=5{,}970$ with the graduation outcome observed; $7.2\%$ attend a Catholic high
school, $36\%$ have Catholic parents; controls are sex, race, parental education and
family income). The outcome is $93\%$ ones. The 2SLS coefficient is $0.153$ (robust
standard error $0.038$), with a first stage of $0.176$ and a reduced form of $0.027$.
The three probit estimators agree with each other at about half the LPM coefficient:
the bivariate-probit APE, a model-based population contrast under the
maintained joint-normal specification, is $0.073$ ($0.007$); the control-function probit with a probit first
stage gives $0.075$ ($0.010$); the linear first stage gives $0.085$ ($0.010$). At
$\gamma=0$ the point estimates of the first two equal \texttt{biprobit} and \texttt{cfprobit}
followed by \texttt{margins}; the control-function standard error adds the direct term of
Proposition~\ref{prop:delta}. The estimated error correlation is
$\hat\rho=-0.40$, negative selection into Catholic schools once a Catholic
background is controlled.

How large would a direct effect of Catholic background on graduation have to be for
the effect to lose significance or sign? For the LPM the answers are $0.014$ and
$0.027$, the second being the reduced form. For the probits they are
$0.019$--$0.024$ and $0.029$--$0.034$, between $7\%$ and $72\%$ larger than the
LPM's: the probit baselines are half the LPM's, but their standard errors are smaller
still and the refit moves them more slowly. The \citet{van_kippersluis_beyond_2018}
direct effect of $0.036$ exceeds all of these tipping values, so it lies beyond the reported
thresholds, provided it can be transported to this sample and calibration. Nor does the analysis
turn into evidence of harm. Their elicited probability-scale prior has variance
$\Omega=(0.125\sqrt{S_0^2+S_{-0}^2})^2$ with $S_0=0.008$ and $S_{-0}=0.012$, so that
$p\sim N(0.036,\,0.0018^{2})$. For the bivariate probit we analyze the index-offset
prior $\gamma_0\sim N(0.310,\,0.0148^{2})$: its mean is the offset whose exact probability
effect is $0.036$ and its standard deviation is $0.0018$ divided by the slope of the
exact map there. Under this plug-in prior, the LTZ interval for the bivariate-probit APE
is $[-0.30,\,0.20]$; the exact probability effects at its $1{,}000$ refits have mean
$0.0361$ and standard deviation $0.0018$, and range from $0.030$ to $0.043$.
The LPM point estimate under the original probability-scale prior becomes
$-0.052$ ($0.040$). The calibration is borrowed rather than re-estimated: their
sample carries controls the public extract lacks, and their reduced form ($0.037$)
almost equals the measured direct effect while ours ($0.027$) does not, so
subtracting the same $0.036$ here overshoots and turns the sign (Online
Appendix~\ref{app:apps}). Across the fitted specifications, a direct effect of
Catholic background of $1.4$ to $2.4$ probability points removes the significance
of the reported contrast and one of $2.7$ to $3.4$ points removes its sign, and the
external benchmark lies beyond both.

\subsection{Angrist--Evans: fertility and female labor supply}
\label{sec:app-ae}

\citet{angrist_children_1998} instrument
having a third child with the first two children being of the same sex and
estimate the effect on whether the mother worked for pay. The instrument is
plausibly random, but the exclusion restriction is contested: a same-sex sibling
pair may change child-rearing costs or parental time use directly, and
\citet{rosenzweig_natural_2000} show that excluding it requires strong
assumptions on preferences and household technology;
\citet{van_kippersluis_beyond_2018} use this design, on Dutch register
data, as their second plausibly-exogenous example. We rebuild the 1980 Census
extract from the authors' public replication files and reproduce their sample
($n=394{,}840$ against their $394{,}835$), first stage ($0.061$) and 2SLS estimate
($-0.117$, standard error $0.025$, against their $-0.120$ and $0.025$).

The probit estimators tell a different story at $\gamma=0$. The bivariate-probit
APE is $-0.211$ ($0.020$) and the control-function probit with a probit first stage
gives $-0.213$ ($0.021$), nearly twice the 2SLS coefficient and with a smaller
standard error; the control-function probit with a linear first stage, whose
first-stage residual is the LPM residual, tracks 2SLS at $-0.117$ ($0.025$).
This numerical agreement supplies no causal interpretation for that fitted
contrast. Differences in estimands and model specifications both matter.
Under joint normality the
bivariate probit delivers an average treatment effect for the population, whereas, under the requisite IV and monotonicity assumptions, a complier
interpretation concerns mothers whose fertility the same-sex pair moved
\citep{angrist_estimation_2001}. A further consideration is identification of the error
correlation: with the same-sex instrument alone the bivariate probit estimates
$\hat\rho=0.059$ (standard error $0.035$), and adding the twin-birth instrument of
the original paper moves the bivariate-probit APE to $-0.126$ ($0.011$) and
$\hat\rho$ to $-0.083$, while 2SLS moves only from $-0.117$ to $-0.090$. The
bivariate-probit APE is sensitive to the instrument set, whether because one instrument
pins the joint-normal $\rho$ loosely or because the two instrument sets are not compatible
with one model, and the
$-0.21$ should not be read as \emph{the} average treatment effect. A sensitivity
comparison therefore holds the estimand and the instrument set fixed, row by row.

Within each row the reading is clean, and only one direction of violation matters: a
same-sex pair that raised the mother's employment on its own would make the 2SLS
estimate an understatement, which is why the box is $[-0.03,\,0]$. For the LPM,
significance is lost at a negative direct effect of $0.4$ probability points and the
sign at $0.7$, the reduced form: a same-sex pair that lowered the mother's employment
probability by seven tenths of a point on its own would account for the whole 2SLS
estimate. For the bivariate and probit-first-stage control-function APEs the
corresponding values are $1.4$--$1.6$ and $1.7$--$1.9$ points, between two and four
times further out, because those baselines are larger and more precisely estimated.
Is a negative direct effect of that size plausible? The closest direct measurement we
know of is the Dutch analysis of \citet{van_kippersluis_beyond_2018}, with two boys as
the instrument and mothers born in countries with a strong son preference as the
zero-first-stage group: the direct effect on employment is $0.001$ (standard error
$0.008$), centered on zero, but its $95\%$ interval runs from $-1.5$ to $1.7$ points.
It spans the LPM's tipping values with room to spare, and its lower end falls between
the significance tipping values of the two probits ($1.4$ and $1.6$). A different
instrument in another country three decades later is not a calibration for the 1980
Census, but the evidence available cannot exclude a violation large enough to matter
for the LPM (Online Appendix~\ref{app:apps} gives the LTZ and two-instrument
results).

\subsection{Confucian clans and bank borrowing}
\label{sec:app-clan}

\citet{chen_banking_2022} argue that the Confucian clan and financial markets are
competing substitutes: regions with historically stronger clans established fewer
modern banks in the four decades after China's first modern bank was founded in
1897, and the clan still limits financial development today. Part of the evidence
for the second claim is at the individual level. In the 2010 Chinese General Social
Survey they relate whether a respondent borrows money from banks to the log number of
clan genealogies per $10{,}000$ inhabitants of the prefecture, instrumented by the log
distance to the nearest of the three academies at which the Neo-Confucian master Zhu
Xi taught, with individual and prefectural controls and clustering by prefecture
(their Table~9, column~2; Online Appendix~\ref{app:apps} lists the controls). They
argue that the academies' locations were set by Zhu Xi's mourning obligations and
official appointments rather than by economic conditions, and support the exclusion
restriction with randomization tests that place counterfactual academies elsewhere.
Distance to an academy could still matter for borrowing through channels other than
clan strength, such as education, geography and attitudes to formal finance, to the
extent the controls do not absorb them.

We use their replication package and reproduce their estimate exactly, an
instrumental-variable probit coefficient of $-0.186$ (standard error $0.078$) on
$n=9{,}074$ respondents in $68$ prefectures, of whom $13.7\%$ borrow from banks. The
cluster-robust first-stage $F$ is $19.4$ and the partial $R^2$ is $0.25$. The 2SLS coefficient
on log clan density is $-0.039$ (standard error $0.017$) and the control-function probit APE is
$-0.039$ ($0.017$), a point estimate equal to \texttt{cfprobit} with \texttt{margins} and to the
average structural marginal effect after their \texttt{ivprobit}; the refit curves in Figure~\ref{fig:apps} lie
almost on top of one another.

Here $\gamma$ is the direct effect of a one-unit change in log distance on the
borrowing probability; the standard deviation of log distance is $0.63$. The
direction that threatens the result is positive: respondents further from an
academy borrowing more for reasons other than weaker clans. Significance is lost
at a direct effect of $0.69$ points per unit under both the LPM and the probit, that
is $0.43$ points per standard deviation; the sign is lost
at $4.39$ and $4.43$ points per unit, $2.75$ and $2.78$ per standard deviation.
The reduced form is $2.75$ points per standard deviation, so a direct channel worth
about $15\%$ of it removes the significance of the clan effect, while reversing its
sign needs a direct channel as large as the whole reduced form. This is the clearest
case in the paper of fragile significance on a stable sign: the conclusion that
stronger clans go with less bank borrowing survives any direct effect smaller than
the reduced form, while its significance survives only a direct effect below half a
point per standard deviation, about $3\%$ of the outcome mean.

\subsection{What the applications show}
\label{sec:app-lessons}

Three regularities run across the applications. First, on the probability scale the
probit and the LPM sensitivity analyses can be compared after aligning
the calibration and estimand: the borrowed Catholic-schooling benchmark exceeds each reported tipping value, the two clan analyses coincide, and in the fertility data the estimators
differ across estimands and maintained specifications. Second, the two tipping values separate two
kinds of fragility: in the clan data significance goes at a violation about $15\%$
of the one that reverses the sign, a fragile significance on a stable sign, while in
the Catholic-schooling and fertility data the two values are within a factor of two.
Third, for the single-instrument LPM the sign-flip tipping value \emph{is} the reduced form, so a
linear sensitivity analysis of the sign adds nothing to a reduced-form table; the
probit sign-flip values range from the reduced form itself (the linear-first-stage
control-function probit in the fertility data) to about $170\%$ above it, and carry
information about the fitted effect under the chosen specification.

\section{Conclusion}
\label{sec:conc}

We specialize the plausibly-exogenous framework to probability-scale effects
in a control-function probit, using the generic nonlinear union and local-to-zero
constructions of the CHR working-paper appendix. The effect variance accounts
for both parameter-mediated and direct first-stage uncertainty; a coefficient-level
Murphy--Topel correction alone is insufficient.

Under the maintained model, a continuum union of valid candidate intervals gives pointwise
asymptotic coverage for a true violation in the stated support, and the LTZ interval gives
prior-averaged coverage along a local sequence; at a fixed prior scale it additionally needs
uniformity and a strictly monotone scalar refit map. The Gaussian linear experiment yields the calibration
plateau $2\Phinorm(\zq c)-1$, not a universal finite-sample formula.

Structural and conditional-index units differ by $\tilde\sigma=\sqrt{1-\rho^2}$
in the linear Gaussian model, and feasible structural calibration must account for the
candidate-specific loading. Probability-scale elicitation is more directly interpretable but
enters through a first-order conversion. Common index violations need not cancel from
interaction effects.

The applications illustrate how significance and sign can respond differently
to departures from exclusion. Binary, fractional and count treatments require separate
structural or control-sufficiency assumptions, which the continuous Gaussian theory does not
by itself validate. Stata and R implementations accompany the analysis.

\bibliographystyle{apalike}
\bibliography{refs}

\begin{thebibliography}{}

\bibitem[Ai and Norton, 2003]{ai_interaction_2003}
Ai, C. and Norton, E.~C. (2003).
\newblock Interaction terms in logit and probit models.
\newblock {\em Economics Letters}, 80(1):123--129.

\bibitem[Altonji et~al., 2005a]{altonji_evaluation_2005}
Altonji, J.~G., Elder, T.~E., and Taber, C.~R. (2005a).
\newblock An evaluation of instrumental variable strategies for estimating the effects of {Catholic} schooling.
\newblock {\em The Journal of Human Resources}, 40(4):791--821.

\bibitem[Altonji et~al., 2005b]{altonji_selection_2005}
Altonji, J.~G., Elder, T.~E., and Taber, C.~R. (2005b).
\newblock Selection on observed and unobserved variables: assessing the effectiveness of {Catholic} schools.
\newblock {\em Journal of Political Economy}, 113(1):151--184.

\bibitem[Angrist, 2001]{angrist_estimation_2001}
Angrist, J.~D. (2001).
\newblock Estimation of limited dependent variable models with dummy endogenous regressors: simple strategies for empirical practice.
\newblock {\em Journal of Business \& Economic Statistics}, 19(1):2--16.

\bibitem[Angrist and Evans, 1998]{angrist_children_1998}
Angrist, J.~D. and Evans, W.~N. (1998).
\newblock Children and their parents' labor supply: evidence from exogenous variation in family size.
\newblock {\em American Economic Review}, 88(3):450--477.

\bibitem[Blundell and Powell, 2003]{blundell_endogeneity_2003}
Blundell, R. and Powell, J.~L. (2003).
\newblock Endogeneity in nonparametric and semiparametric regression models.
\newblock In Dewatripont, M., Hansen, L.~P., and Turnovsky, S.~J., editors, {\em Advances in Economics and Econometrics: Theory and Applications, Eighth World Congress}, volume~2 of {\em Econometric Society Monographs}, pages 312--357. Cambridge University Press, Cambridge.

\bibitem[Blundell and Powell, 2004]{blundell_endogeneity_2004}
Blundell, R.~W. and Powell, J.~L. (2004).
\newblock Endogeneity in {Semiparametric} {Binary} {Response} {Models}.
\newblock {\em Review of Economic Studies}, 71(3):655--679.

\bibitem[Chen et~al., 2023]{chen_another_2023}
Chen, K., Martin, R.~S., and Wooldridge, J.~M. (2023).
\newblock Another {Look} at the {Linear} {Probability} {Model} and {Nonlinear} {Index} {Models}.
\newblock arXiv:2308.15338.

\bibitem[Chen et~al., 2022]{chen_banking_2022}
Chen, Z., Ma, C., and Sinclair, A.~J. (2022).
\newblock Banking on the {Confucian} clan: why {China} developed financial markets so late.
\newblock {\em The Economic Journal}, 132(644):1378--1413.

\bibitem[Clarke and Matta, 2018]{clarke_practical_2018}
Clarke, D. and Matta, B. (2018).
\newblock Practical {Considerations} for {Questionable} {IVs}.
\newblock {\em The Stata Journal}, 18(3):663--691.

\bibitem[Conley et~al., 2008]{conley_plausibly_2008}
Conley, T.~G., Hansen, C.~B., and Rossi, P.~E. (2008).
\newblock Plausibly {Exogenous}.
\newblock Working paper, Graduate School of Business, University of Chicago; draft of 31 July 2008.

\bibitem[Conley et~al., 2012]{conley_plausibly_2012}
Conley, T.~G., Hansen, C.~B., and Rossi, P.~E. (2012).
\newblock Plausibly {Exogenous}.
\newblock {\em Review of Economics and Statistics}, 94(1):260--272.

\bibitem[Greene, 2010]{greene_testing_2010}
Greene, W. (2010).
\newblock Testing hypotheses about interaction terms in nonlinear models.
\newblock {\em Economics Letters}, 107(2):291--296.

\bibitem[Han and Lee, 2019]{han_estimation_2019}
Han, S. and Lee, S. (2019).
\newblock Estimation in a generalization of bivariate probit models with dummy endogenous regressors.
\newblock {\em Journal of Applied Econometrics}, 34(6):994--1015.

\bibitem[Kraay, 2012]{kraay_instrumental_2012}
Kraay, A. (2012).
\newblock Instrumental variables regressions with uncertain exclusion restrictions: a {Bayesian} approach.
\newblock {\em Journal of Applied Econometrics}, 27(1):108--128.

\bibitem[Murphy and Topel, 1985]{murphy_estimation_1985}
Murphy, K.~M. and Topel, R.~H. (1985).
\newblock Estimation and {Inference} in {Two}-{Step} {Econometric} {Models}.
\newblock {\em Journal of Business \& Economic Statistics}, 3(4):370--379.

\bibitem[Nevo and Rosen, 2012]{nevo_identification_2012}
Nevo, A. and Rosen, A.~M. (2012).
\newblock Identification {With} {Imperfect} {Instruments}.
\newblock {\em Review of Economics and Statistics}, 94(3):659--671.

\bibitem[Newey, 1985]{newey_generalized_1985}
Newey, W.~K. (1985).
\newblock Generalized method of moments specification testing.
\newblock {\em Journal of Econometrics}, 29(3):229--256.

\bibitem[Newey and McFadden, 1994]{newey_large_1994}
Newey, W.~K. and McFadden, D. (1994).
\newblock Large sample estimation and hypothesis testing.
\newblock In {\em Handbook of {Econometrics}}, volume~4, pages 2111--2245. Elsevier.

\bibitem[Pagan, 1984]{pagan_econometric_1984}
Pagan, A. (1984).
\newblock Econometric {Issues} in the {Analysis} of {Regressions} with {Generated} {Regressors}.
\newblock {\em International Economic Review}, 25(1):221--247.

\bibitem[Puhani, 2012]{puhani_treatment_2012}
Puhani, P.~A. (2012).
\newblock The treatment effect, the cross difference, and the interaction term in nonlinear “difference-in-differences” models.
\newblock {\em Economics Letters}, 115(1):85--87.

\bibitem[Radean, 2023]{radean_ginteff_2023}
Radean, M. (2023).
\newblock ginteff: {A} generalized command for computing interaction effects.
\newblock {\em The Stata Journal}, 23(2):301--335.

\bibitem[Rivers and Vuong, 1988]{rivers_limited_1988}
Rivers, D. and Vuong, Q.~H. (1988).
\newblock Limited information estimators and exogeneity tests for simultaneous probit models.
\newblock {\em Journal of Econometrics}, 39(3):347--366.

\bibitem[Rosenzweig and Wolpin, 2000]{rosenzweig_natural_2000}
Rosenzweig, M.~R. and Wolpin, K.~I. (2000).
\newblock Natural ``natural experiments'' in economics.
\newblock {\em Journal of Economic Literature}, 38(4):827--874.

\bibitem[Terza et~al., 2008]{terza_two_stage_2008}
Terza, J.~V., Basu, A., and Rathouz, P.~J. (2008).
\newblock Two-stage residual inclusion estimation: {Addressing} endogeneity in health econometric modeling.
\newblock {\em Journal of Health Economics}, 27(3):531--543.

\bibitem[Van~Kippersluis and Rietveld, 2018]{van_kippersluis_beyond_2018}
Van~Kippersluis, H. and Rietveld, C.~A. (2018).
\newblock Beyond plausibly exogenous.
\newblock {\em The Econometrics Journal}, 21(3):316--331.

\bibitem[Wooldridge, 2005]{wooldridge_unobserved_2005}
Wooldridge, J.~M. (2005).
\newblock Unobserved {Heterogeneity} and {Estimation} of {Average} {Partial} {Effects}.
\newblock In Andrews, D. W.~K. and Stock, J.~H., editors, {\em Identification and {Inference} for {Econometric} {Models}}, pages 27--55. Cambridge University Press, 1 edition.

\bibitem[Wooldridge, 2014]{wooldridge_quasi_maximum_2014}
Wooldridge, J.~M. (2014).
\newblock Quasi-maximum likelihood estimation and testing for nonlinear models with endogenous explanatory variables.
\newblock {\em Journal of Econometrics}, 182(1):226--234.

\bibitem[Wooldridge, 2015]{wooldridge_control_2015}
Wooldridge, J.~M. (2015).
\newblock Control {Function} {Methods} in {Applied} {Econometrics}.
\newblock {\em Journal of Human Resources}, 50(2):420--445.

\end{thebibliography}

\clearpage
\begin{center}
{\Large\textbf{Online Appendix}}
\end{center}

\noindent Sections, equations, propositions, tables and figures numbered without a letter
refer to the main article; those with a letter prefix are in this Online Appendix.

\appendix
\counterwithin{equation}{section}
\counterwithin{table}{section}
\counterwithin{figure}{section}

\section{Regularity conditions and proofs}
\label{app:proofs}

\subsection{Conditions}
\label{app:cond}

\begin{itemize}[leftmargin=2.2em,nosep]
\item[\textbf{R1}] $\{(y_i,d_i,x_i,z_i)\}_{i=1}^n$ is i.i.d.\ with
$\E\|w_i\|^4<\infty$, $\E d_i^4<\infty$.
\item[\textbf{R2}] Assumption~\ref{a:cf} holds (linear first stage), or, for a
nonlinear first stage, the generalized-residual control function of
Appendix~\ref{app:firststage} is the maintained model: $\varepsilon_i$ depends on
$(w_i,d_i)$ only through the control function and the response probability has the
probit form in it, an assumption distinct from, and not nested in,
Assumption~\ref{a:cf}.
\item[\textbf{R3}] $\E[w_iw_i\trans]$ is nonsingular and $\pi_z\ne0$ (instrument
relevance); the augmented probit information
$\E[\omega(\eta_i)g_ig_i\trans]$ is nonsingular at the true parameters; the violation
lies in a bounded set.
\item[\textbf{R4}] The maps $(\vartheta,\gamma_0)\mapsto q(W,\vartheta,\gamma_0)$ and the
probit objective are twice continuously differentiable with envelopes possessing
finite second moments on a neighborhood of the truth; the pushforward distributions
appearing in Proposition~\ref{prop:plateau} are continuous with positive density at
the relevant quantiles. (Smoothness of $\Phinorm,\phinorm$ makes the first part
automatic; the envelope conditions follow from R1.)
\item[\textbf{R5}] $(\hat\pi,\hat\vartheta)$ are consistent for interior
$(\pi,\vartheta)$ (standard primitive conditions in
\citealp[Thms.~2.6, 6.1]{newey_large_1994}).
\item[\textbf{R6}] $\gamma_n=\nu/\sqrt n$ with $\nu\sim F$, $\E\|\nu\|^2<\infty$, and
$\nu$ drawn independently of the covariates and disturbances
$\{(x_i,z_i,\varepsilon_i,v_i)\}$. The outcomes are then generated under $\gamma_n$, so
they depend on $\nu$. The influence expansion behind
Propositions~\ref{prop:normal}--\ref{prop:delta} and the central limit theorem hold
conditionally on $\nu$, uniformly over each bounded set of $\nu$, with the same
$N(0,V_\theta)$ limit.
\end{itemize}

Proofs of Propositions~\ref{prop:normal}--\ref{prop:uci} are in Appendix~\ref{app:proofs13}. Throughout,
$m(\gamma_0;\gamma)$ is the pseudo-true map \eqref{eq:map}; under R1--R5 it is
continuously differentiable near the origin by the implicit function theorem applied
to $\E[s(\vartheta;\gamma_0,\gamma)]=0$ (the expected score is smooth in all three
arguments by R4), and differentiating the identity $m(\gamma;\gamma)=\theta(\gamma)$
gives
\begin{equation}
\partial_1 m(0;0) + \partial_2 m(0;0) = \theta'(0), \qquad A \equiv -\partial_1 m(0;0).
\label{eq:chain}
\end{equation}

\emph{Closed form of the cross-derivative $G$ (linear first stage).} Write
$s_i=\lambda(\eta_i)(y_i-\Phinorm(\eta_i))g_i$ and note that $\hat\pi$ enters through
the index, $\partial\eta_i/\partial\pi\trans=-\tcf w_i\trans$, and through the
$\hat v_i$ entry of $g_i$, $\partial g_i/\partial\pi\trans = -e_{v}w_i\trans$ with
$e_{v}$ the selection vector of the control-function row. At the true parameters,
$\E[\partial\{\lambda(\eta)(y-\Phinorm(\eta))\}/\partial\eta \given
x,d,z,v]=-\omega(\eta)$, while the second channel has conditional expectation
proportional to $\E[\lambda(\eta)(y-\Phinorm(\eta))\given x,d,z,v]=0$. Hence
\[
G \;=\; \tcf\,\E\big[\omega(\eta_i)\, g_i w_i\trans\big].
\]
For a nonlinear first stage the same computation applies with
$\partial \hat v_i/\partial\pi\trans$ given by the generalized-residual derivative,
which introduces the per-observation weights of Appendix~B.

\subsection{Proofs of Propositions~\ref{prop:normal}--\ref{prop:uci}}
\label{app:proofs13}

\begin{proof}[Proof of Proposition~\ref{prop:normal}]
By R5, $\hat\vartheta\xrightarrow{p}\vartheta$ and a mean-value expansion of the second-stage
moment about $(\vartheta,\pi)$ gives
\[
0 = n^{-1}\!\sum_i s_i(\vartheta,\pi) + \bar H_n(\hat\vartheta-\vartheta) + \bar G_n(\hat\pi-\pi) + o_p(n^{-1/2}),
\]
where $\bar H_n, \bar G_n$ converge in probability to $H,G$ of \eqref{eq:HG} by R4 and a
uniform law of large numbers. Solving and substituting the first-stage expansion
$\sqrt n(\hat\pi-\pi)=n^{-1/2}\sum_i\psi_{1i}+o_p(1)$,
\[
\sqrt n(\hat\vartheta-\vartheta) = -H^{-1}\, n^{-1/2}\!\sum_i\big(s_i(\vartheta,\pi)+G\psi_{1i}\big) + o_p(1).
\]
The summands are i.i.d.\ with mean zero (the probit score is a martingale-difference at
the truth and $\E[\psi_{1i}]=0$) and finite variance by R1--R4; the Lindeberg--L\'evy CLT
and the continuous-mapping theorem give the stated limit with
$V_\vartheta = H^{-1}\Sigma H^{-\prime}$. Expanding $\Sigma$ and using
$V_1=\E[\psi_{1i}\psi_{1i}\trans]$, $C=\E[s_i\psi_{1i}\trans]$ yields the
\citet{murphy_estimation_1985} form. This is the two-step M-estimator argument of
\citet[Thm.~6.1]{newey_large_1994}, specialized to the probit second stage with an OLS
first stage; the generated regressor contributes through $G$
\citep{pagan_econometric_1984}.
\end{proof}

\emph{Computation.} $H$, $G$, $S$, $C$, and $V_1$ are sample averages of known functions of
$(\hat\vartheta,\hat\pi)$ and the data; $\hat V_\vartheta$ is assembled in closed form without
resampling. For a linear first stage the construction is the Murphy--Topel sandwich
implemented in \texttt{apexog}. The APE standard error built from it alone, the first term
of \eqref{eq:if-sample}, agrees with \texttt{cfprobit} followed by \texttt{margins} to
nine decimals at $\gamma=0$, and it reproduces the generated-regressor inflation of the
naive probit standard error elsewhere (Section~\ref{sec:mc} and Appendix~C).

\begin{proof}[Proof of Proposition~\ref{prop:delta}]
A joint Taylor expansion of the sample average of $q$ gives
\[
\sqrt n(\hat\theta-\theta_n)
 =J_\vartheta\trans\sqrt n(\hat\vartheta-\vartheta)
  +J_\pi\trans\sqrt n(\hat\pi-\pi)+o_p(1).
\]
Substituting the two parameter influence functions proves \eqref{eq:if-sample}. Since
$\E[\psi_{2i}\psi_{1i}\trans]=-H^{-1}(C+GV_1)$, expanding its variance gives
\eqref{eq:full-effect-var}. For a population target add
$\sqrt n(\theta_n-\theta)=n^{-1/2}\sum_i(q_i-\theta)$ before taking the variance. The joint
central limit theorem and consistent sample derivatives complete the argument.
\end{proof}

\begin{proof}[Proof of Proposition~\ref{prop:uci}]
At $\gamma_0=\gamma$ the offset equals the true violation, the model in
\eqref{eq:cfprobit} is correctly specified, and Proposition~\ref{prop:delta} gives
$\Prob\{\theta_n(\gamma)\in\mathrm{CI}_n(\gamma)\}\to 1-\alpha$. Since $\gamma\in\calG$,
$\mathrm{UCI}_n \supseteq \mathrm{CI}_n(\gamma)$ by construction, and coverage is
monotone under set inclusion. This is the union construction of
\citet[\S 4.1]{conley_plausibly_2008}, stated there for the 2SLS coefficient and
extended to generic M-estimators in their \S 8.7, specialized to the two-step
APE/AIE; the content of the proposition is that Proposition~\ref{prop:delta}
supplies the per-$\gamma_0$ variance estimator that construction assumes.
\end{proof}

\emph{The finite grid.} The software approximates the extrema of the union on a finite grid,
and inclusion of the truth in the box does not imply that the grid contains its Wald interval.
A sufficient condition for preserving the coverage result is an $o_p(n^{-1/2})$ error in both
extrema (with nondegenerate root-$n$ uncertainty), or conservative outward enlargement by
certified error bounds. For Lipschitz endpoints with $O_p(1)$ Lipschitz constants, mesh size
$o(n^{-1/2})$ suffices. A fixed coarse grid alone has no general continuum-coverage guarantee;
when the endpoints are monotone in $\gamma_0$ the extrema sit at the corners of the box and any
grid containing them is exact.

\subsection{Proof of Proposition~\ref{prop:ltz}}
\label{app:proof-ltz}

Condition on $\nu$. The data are generated under $\gamma_n=\nu/\sqrt n\to0$, a
drifting sequence of DGPs converging to the $\gamma=0$ model. Let
$\vartheta^*(\gamma_0;\gamma)$ be the pseudo-true second-stage parameter at assumed
offset $\gamma_0$ under true violation $\gamma$, and let
$m_n(\gamma_0;\gamma)\equiv n^{-1}\sum_i q(W_i,\vartheta^*(\gamma_0;\gamma),\gamma_0)$
be the sample analog of the map \eqref{eq:map}, so that
$m_n(\gamma;\gamma)=\theta_n(\gamma)$ and $m_n\to m$ together with its first
derivatives uniformly on compacta (R4 envelopes). Decompose
\[
\sqrt n\big(\hat\theta(0)-\theta_n(\gamma_n)\big)
= \underbrace{\sqrt n\big(\hat\theta(0)-m_n(0;\gamma_n)\big)}_{(\mathrm{I})}
+ \underbrace{\sqrt n\big(m_n(0;\gamma_n)-m_n(\gamma_n;\gamma_n)\big)}_{(\mathrm{II})}.
\]
For (I): $\hat\theta(0)-m_n(0;\gamma_n)$ is the plug-in functional's deviation from
its value at the pseudo-true parameter, holding the sample covariates fixed; the
expansion in the proofs of Propositions~\ref{prop:normal}--\ref{prop:delta} holds
uniformly over $\gamma$ in $n^{-1/2}$-neighborhoods of $0$ by R4 (the score,
Hessian, and cross-derivative are continuous in the DGP parameter, and the
Lindeberg condition holds along the triangular array), so
$(\mathrm{I})\xrightarrow{d}N(0,V_\theta)$ with $V_\theta$ evaluated at $\gamma=0$.
For (II): by the mean-value theorem and continuous differentiability of $m_n$,
$(\mathrm{II}) = -\partial_1 m_n(0;0)\,\sqrt n\,\gamma_n + o_p(1) = A\nu + o_p(1)$,
since $\partial_1 m_n(0;0)\to\partial_1 m(0;0)=-A$ by \eqref{eq:chain} and the
uniform convergence of the derivatives. Both statements hold uniformly over
$\|\nu\|\le M$ for each fixed $M$ (R6), and the conditional limit law of (I) does not
depend on $\nu$. Hence, for bounded continuous $g$ and $h$ and $Z\sim N(0,V_\theta)$,
$\E[g(\mathrm{I})\,h(\nu)\,\indic\{\|\nu\|\le M\}]\to
\E[g(Z)]\,\E[h(\nu)\,\indic\{\|\nu\|\le M\}]$ by dominated convergence, and the
contribution of $\|\nu\|>M$ is at most $\sup|g|\sup|h|\,F\{\|\nu\|>M\}$, which vanishes as
$M\to\infty$. So $(\mathrm{I},\nu)$ converges jointly to $(Z,\nu)$ with $Z$ independent of
$\nu$, although in any finite sample the outcomes depend on $\nu$; the continuous mapping
theorem then gives the independent sum.
\hfill$\square$

\emph{Proof of Corollary~\ref{cor:exact}.} Write $\gamma^{(j)}=\nu^{(j)}/\sqrt n$,
$\nu^{(j)}\sim F$. On each fixed set $\|\nu^{(j)}\|\le M$, the local uniform expansion
(now in the assumed argument) and consistent variance estimation approximate the centered
draws $\sqrt n\{\theta^{*(j)}-\hat\theta(0)\}$ by
$-A\nu^{(j)}+\sqrt{V_\theta}\,\epsilon_j$ in conditional distribution. No expansion is
claimed uniformly over all draws: the fraction of prior draws outside this set converges to
$F\{\|\nu\|>M\}$ as $B\to\infty$, and letting $M\to\infty$ therefore establishes
convergence of the full conditional draw distribution to that of $Z^*-A\nu^*$, where
$Z^*\sim N(0,V_\theta)$ and $\nu^*\sim F$ are independent. The endpoint quantiles
consequently converge for any $B=B_n\to\infty$, with no rate linking $B$ and $n$. By
symmetry of the normal component, the limiting draw distribution is that of the negative
of the limiting pivot in Proposition~\ref{prop:ltz}, so the percentile interval has
prior-averaged coverage $1-\alpha$. Only the symmetry of the Gaussian noise is used, not
that of $F$: because each refit subtracts the assumed violation, the percentile interval
of the convolved refits is the pivot interval of \citet[\S 4.3]{conley_plausibly_2008}
for \emph{any} prior, symmetric or not.
\hfill$\square$

\subsection{Proof of Proposition~\ref{prop:plateau}}
\label{app:proof-plateau}

\emph{(a)} Condition on $\gamma$ (a draw from $F$, fixed as $n\to\infty$). The
statement assumes compactly supported $F$; an unbounded prior such as the Gaussian
is handled by truncation, since for any $\epsilon>0$ a compact set carries mass
$1-\epsilon$ under $F$, the argument below applies on that set, and the coverage
contribution of the tails is at most $\epsilon$; letting $\epsilon\to0$ delivers the
Gaussian case (the truncation must respect monotonicity of the map on the enlarged
set, which R4's differentiability and the maintained monotonicity provide on
compacta). By R1--R5
and a uniform law of large numbers over the compact support of $F$ (R4 envelopes),
$\sup_{\gamma_0\in\mathrm{supp}F}|\hat\theta(\gamma_0)-m(\gamma_0;\gamma)|
\xrightarrow{p}0$ and
$\sup_{\gamma_0}\hat\sigma(\gamma_0)=O_p(n^{-1/2})\to0$. Hence the convolved draws
$\theta^{*(j)}=\hat\theta(\gamma^{(j)})+\hat\sigma(\gamma^{(j)})\epsilon_j$, with
$\gamma^{(j)}\sim F$ i.i.d., have empirical distribution converging weakly to the
pushforward $m(\cdot\,;\gamma)_{\#}F$ as $n$ and the number of draws grow; its
$\alpha/2$ and $1-\alpha/2$ empirical quantiles converge to those of the pushforward
(continuity and positive density, R4). With $m(\cdot\,;\gamma)$ strictly increasing
(the decreasing case swaps the endpoints), monotone quantile equivariance gives
$Q_p(\gamma) = m\big(Q^F_p;\gamma\big)$, so the coverage event for the target
$\theta(\gamma)=m(\gamma;\gamma)$ is
\[
m\big(Q^F_{\alpha/2};\gamma\big) \;\le\; m(\gamma;\gamma) \;\le\;
m\big(Q^F_{1-\alpha/2};\gamma\big)
\;\Longleftrightarrow\;
Q^F_{\alpha/2} \;\le\; \gamma \;\le\; Q^F_{1-\alpha/2},
\]
an event of probability exactly $1-\alpha$ under continuous $F$. Note what was
\emph{not} used: no linearity, no separability, no restriction on how $m$ varies with
its second argument. The target is the map evaluated at the true draw, so
monotonicity in the first argument alone reduces the event to ``$\gamma$ inside its
own quantiles.'' Dominated convergence over $\gamma\sim F$ completes the limit.

\emph{(b)} Work in the Gaussian limit experiment implied by
Proposition~\ref{prop:normal}: $\hat\theta(0)-\theta_n(\gamma) = A\gamma + \xi$ with
$\xi\sim N(0,V_\theta/n)$, $\gamma\sim N(0,\sigma_{\text{true}}^2)$ independent, and
$\sigma_{\text{true}}=\sigma_{\text{stated}}/c$; the finite-sample ratio $\varrho$ is
held fixed, so the statement is exact in the limit experiment and an approximation
at finite $n$. With the map linear in $\gamma_0$,
the refits satisfy $\hat\theta(\gamma_0)=\hat\theta(0)-A\gamma_0$, so the convolved
draws are $\hat\theta(0) - A\sigma_{\text{stated}}U_j + \hat\sigma(\gamma^{(j)})\epsilon_j$, with
$U_j\sim N(0,1)$ i.i.d.\ standardized draws. Under the assumed uniform consistency
of $\hat\sigma(\cdot)$ over the stated support, the per-draw scales may be replaced
by the common $\sqrt{V_\theta/n}$ at the cost of an $o_p(1)$ perturbation of the
quantiles, and,
as the number of draws grows, the interval converges to
$\hat\theta(0)\pm \zq\sqrt{A^2\sigma_{\text{stated}}^2+V_\theta/n}$
(Gaussian convolution). Coverage is then
\[
\Prob\Big\{\big|N\big(0,\,A^2\sigma_{\text{stated}}^2/c^2+V_\theta/n\big)\big|
\le \zq\sqrt{A^2\sigma_{\text{stated}}^2+V_\theta/n}\Big\}
= 2\Phinorm\!\left(\zq\frac{\sqrt{1+\varrho^2}}{\sqrt{1+\varrho^2/c^2}}\right)-1,
\]
dividing through by $V_\theta/n$ and using
$\varrho=|A|\sigma_{\text{stated}}/\sqrt{V_\theta/n}$. Monotonicity in $\varrho$ and
the limits at $\varrho\in\{0,\infty\}$ are immediate.

\emph{(c)} The prior-dominated limit is $n\to\infty$ at fixed prior scale, which
drives the sampling terms to zero and $\varrho\to\infty$. By the
argument of part (a) with the \emph{stated} prior generating the draws, the event
becomes $\gamma \in
[\,Q^{F_{\text{stated}}}_{\alpha/2},\,Q^{F_{\text{stated}}}_{1-\alpha/2}\,]$ with
$\gamma\sim F_{\text{true}}$. Standardizing, $\gamma=\sigma_{\text{true}}U$ and
$Q^{F_{\text{stated}}}_p = c\,\sigma_{\text{true}}\,Q^{F_1}_p$, so for $F_1$ symmetric
about zero the probability is $\Prob\{|U|\le c\,Q^{F_1}_{1-\alpha/2}\}$, free of every
design quantity; only monotonicity of the map was used. \hfill$\square$

\emph{On Remark~\ref{cor:aie}.} A common additive violation need not cancel from an
interaction effect, so no general immunity holds. Where the prior-induced components of the
interval are negligible against sampling noise, $\varrho\approx0$ in \eqref{eq:kappaform} and
coverage stays near $1-\alpha$ for any calibration $c$; this is an approximation statement in
that regime, and whether a given design is in it is an empirical matter. In the design of
Table~\ref{tab:ltz} the AIE covers at $0.958$--$0.960$ against the $c=0.8$ APE plateau of $0.88$.

\emph{Vector violations.} The scalar-violation restriction matters for practice: with $k_z>1$ instruments the
quantile-equivariance argument requires additional structure on the map
$\mathbb{R}^{k_z}\to\mathbb{R}$, and a general direction-dependent miscalibration is
beyond the present theory. When all components share a common calibration factor
$c$, however, the linear map reduces the draws' effect to a scalar index and the
plateau formula \eqref{eq:plateau} extends along that one-dimensional family. Our
two-instrument cell (Section~\ref{sec:mc-ltz}) lies in the transition and its
overidentified refit map, which we have not measured, need not be exactly linear, so it does
not test this reduction.

\section{First-stage variants, binary treatments, and multiple endogenous regressors}
\label{app:firststage}

\subsection{Nonlinear first stages}
\label{app:nlfs}

For a binary, fractional, or count endogenous regressor the linear first stage is
replaced by the corresponding index model and the control function by its generalized
residual \citep{rivers_limited_1988, terza_two_stage_2008, wooldridge_control_2015,
wooldridge_quasi_maximum_2014}:
\begin{center}
\begin{tabular}{lcc}
\toprule
First stage & Control function $\hat v_i$ & Weight $h_i$ \\
\midrule
Linear (OLS) & $d_i - w_i\trans\hat\pi$ & $1$ \\[4pt]
Probit &
$\big(d_i-\Phinorm(w_i\trans\hat\pi)\big)\lambda(w_i\trans\hat\pi)$ &
$\hat v_i\big(w_i\trans\hat\pi+\hat v_i\big)$ \\[4pt]
Fractional probit &
$\big(d_i-\Phinorm(w_i\trans\hat\pi)\big)\lambda(w_i\trans\hat\pi)$ &
$\hat v_i\big(w_i\trans\hat\pi+\hat v_i\big)+d_i(1-d_i)\lambda(w_i\trans\hat\pi)^2$ \\[4pt]
Poisson & $d_i-\exp(w_i\trans\hat\pi)$ & $\exp(w_i\trans\hat\pi)$ \\
\bottomrule
\end{tabular}
\end{center}
Here $\lambda(\cdot)=\phinorm(\cdot)/[\Phinorm(\cdot)(1-\Phinorm(\cdot))]$ as in
\eqref{eq:score}. The weight $h_i=-\partial\hat v_i/\partial(w_i\trans\pi)$ is the
per-observation curvature of the first-stage moment. The probit entry is an identity for a
$0/1$ regressor only; a fractional regressor adds the last term (versions of \texttt{apexog}
before $0.21.1$ omitted it; because the two places the weight enters largely offset, the mean
standard error in the fractional design moved by at most $3\%$, in either direction). The
weight enters
the Murphy--Topel construction through
$M=\E[h_i w_iw_i\trans]$ (first-stage bread) and through the $\hat v_i$-channel of the
cross-derivative $G$, replacing the linear case's $M=\E[w_iw_i\trans]$ and unit
weights, and through the direct gradient $J_\pi$, where $\partial\hat v_i/\partial\pi=-h_iw_i$. No
other part of Section~\ref{sec:est} changes. Our implementation of each
variant reproduces the corresponding built-in two-step fit at $\gamma=0$ to at least
eleven decimals (Appendix~\ref{app:sims}); for the fractional-probit first stage the
\emph{coefficients} agree exactly and their Murphy--Topel standard errors agree with the
built-in command's to $0.01\%$, while its post-estimation APE
uses a different control-function reconstruction and differs from ours by $0.064$ in the
validation design ($0.172$ against $0.236$), a definitional difference documented in the
replication files; we report the structural-slope APE consistently.

\emph{Maintained assumption for nonlinear first stages.} Assumption~\ref{a:cf}
delivers the probit \eqref{eq:cfprobit} by conditioning on the first-stage error
$v_i$. When $d_i$ is binary, fractional or a count, one conditions on $d_i$ rather
than on $v_i$, and joint normality no longer produces a probit in the generalized
residual: for $d_i=\indic\{w_i\trans\pi+v_i>0\}$,
$\E[\varepsilon_i\given d_i,w_i]=\rho\,\hat v_i$ is linear in the generalized
residual, but $\Var(\varepsilon_i\given d_i,w_i)=1-\rho^2 h_i$ is heteroskedastic
and the conditional law is not normal. The generalized-residual control function is
therefore the two-step approach of
\citet{wooldridge_quasi_maximum_2014, wooldridge_control_2015}, consistent under the
sufficiency assumption that $\varepsilon_i$ depends on $(w_i,d_i)$ only through the
control function, $D(\varepsilon_i\given w_i,d_i)=D(\varepsilon_i\given \hat v_i)$,
together with the probit form for $\Prob(y_i=1\given x_i,d_i,\hat v_i)$ (condition
R2). Wooldridge notes that these assumptions and bivariate normality are different
and that neither set nests the other; under a bivariate-normal data-generating
process the two-step estimator is an approximation whose error grows with $\rho^2$.
In the design of Table~\ref{tab:uci} (binary $d$, probit first stage) at
$n=400{,}000$ the two-step discrete-change APE is below the structural value by
$2.7\%$, $11.6\%$ and $50.6\%$ at $\rho=0.3$, $0.5$ and $0.8$, while the recursive
bivariate probit, the exact likelihood under joint normality, is within $0.4\%$
(replication files). At $n=5{,}000$ and $\rho=0.5$ (Table~\ref{tab:uci}) the same
bias is $-0.024$, or $-11\%$, with Wald coverage $0.881$ at $\gamma=0$ and a
correctly sized variance (ratio of sampling standard deviation to reported standard
error $0.99$--$1.02$), while the bivariate probit at the true offset is unbiased at
every violation ($0.0003$, $0.0000$, $0.0002$). The converse holds under a design
that satisfies the sufficiency assumption exactly: $y_i\sim$
Bernoulli$\big(\Phinorm(x_i\trans\beta+\delta d_i+\gamma z_i+\theta\,\hat v_i)\big)$
with $d_i=\indic\{w_i\trans\pi+v_i>0\}$, $\hat v_i$ the generalized residual at the
true $\pi$, $\theta=0.58$ (the pseudo-true loading of the bivariate-normal design at
$\rho=0.5$) and the same $\beta$, $\delta$ and $\pi$. There the two-step is unbiased
($0.0007$) with Wald coverage $0.951$, $0.954$ and $0.944$ at the three violations,
and the bivariate probit is biased by $+0.025$ ($+13\%$) at $\gamma=0$; the
single-average and double-average discrete changes nearly coincide in that design
($0.197$ and $0.198$). The fractional design of Table~\ref{tab:uci},
$d_i=\Phinorm(w_i\trans\pi+u_i)$, is invertible in $u_i$, so an exact control
variable exists ($\Phinorm^{-1}(d_i)-w_i\trans\pi$, a linear first stage on the
transformed $d$) and the generalized residual approximates it: the two-step is
biased by $-0.022$ ($-8\%$) against the conditional-on-$u$ APE and by $-0.038$
($-14\%$) against the structural slope, with the variance right (ratio $1.02$). In
the Poisson design no control function recovers $u_i$; the two-step is unbiased for
the structural slope APE at $\gamma=0$ ($0.0002$, Wald coverage $0.953$) and $6\%$
above the conditional-on-$u$ oracle. Two consequences follow for the units factor
of Section~\ref{sec:coverage}. First, $\sqrt{1-\rho^2}$ is the index-scale
conversion for the linear first stage only: for the generalized-residual engines
the refit at the unscaled structural $\gamma$ is closer to the truth than the refit
at $\gamma/\sqrt{1-\rho^2}$ (Wald coverage $0.936$ against $0.590$ for binary $d$
at $\gamma=0.2$, $0.949$ against $0.877$ for the fractional probit), interpolating
between the two offsets puts the zero-bias offset within $5\%$ of the structural
$\gamma$, so no conversion should be applied there, and the reported
$\hat{\tilde\sigma}$ ($0.92$, $0.92$ and $0.98$ at $\gamma=0$ in the three designs)
is not that factor. Second, $\hat{\tilde\sigma}$ is a baseline-fit quantity that
moves with the violation ($0.80$, $0.91$ and $1.00$ in the linear design and
$0.98$, $0.93$ and $0.61$ in the Poisson design at $\gamma=0$, $0.2$ and $0.5$),
which bears on \texttt{gunits(structural)} under wide priors. For a
binary treatment the recursive bivariate probit, which accepts the violation as an
offset in the outcome equation and has a standard likelihood variance, is the exact
alternative under joint normality; \texttt{apexog} offers it as
\texttt{firststage(biprobit)} (version 0.19), refitting the recursive bivariate
probit at each violation and reporting the ASF discrete change with a
likelihood-based delta-method variance. The two estimators rest on non-nested
assumptions, so the choice between them is a modeling decision, not a ranking.

\subsection{Binary endogenous treatments}
\label{app:binary}

When $d_i\in\{0,1\}$ the APE switches automatically to the discrete-change form
\eqref{eq:ape-bin}; the probit first stage supplies the generalized residual above,
under the maintained assumption just stated, or the recursive bivariate probit of
the preceding paragraph replaces the control function altogether.
Discrete-change effects propagate the violation more strongly through the nonlinear
index than slope effects do, which widens sensitivity bands: the binary-$d$ unions
of Table~\ref{tab:uci} are $0.78$, $0.73$ and $0.40$ wide against $0.165$, $0.157$
and $0.145$ for the linear first stage.

\subsection{Multiple endogenous regressors and endogenous interactions}
\label{app:multi}

With $K$ endogenous regressors, each has its own first stage (possibly of different
type, possibly with its own instrument subset), its own control function, and the
second-stage index includes all $K$ control functions. The Murphy--Topel correction
sums the $K$ first-stage channels: $G=\sum_k G_k$ with per-equation cross-derivatives,
and the influence function of each reported effect runs over the full parameter vector
and every first stage's direct gradient (Appendix~\ref{app:jac}), so every effect's standard
error absorbs estimation uncertainty from every first stage. The units factor generalizes accordingly: with control-function loading
vector $b_{\mathrm{cf}}$ and control-function covariance $\Var(\hat v)$, the estimated
conversion is $\hat{\tilde\sigma}=(1+b_{\mathrm{cf}}\trans\Var(\hat v)\,b_{\mathrm{cf}})^{-1/2}$,
of which \eqref{eq:sigtilde} is the scalar case; this is what \texttt{gunits(structural)}
computes.

The empirically important special case is an \emph{endogenous interaction}: a
treatment $d$ interacted with a moderator $x_m$, making $d\times x_m$ a second
endogenous regressor. It receives its own first stage on the user-supplied augmented
instrument set (instruments and their moderator interactions) and its own control
function; products of the control functions are added to the index, mirroring the
interacted-control-function specification of the built-in two-step command, and the
cross-derivative $G$ acquires the corresponding product terms. When no interactions
are present the formulas collapse exactly to the single-treatment case. The AIE for
this configuration is the one validated against the \texttt{ginteff} oracle
\citep{radean_ginteff_2023} to eleven decimals and covered at nominal rates in
Table~\ref{tab:ltz}. The product need not be declared endogenous: with the
moderator-interacted instruments in the treatment's first stage, the treatment's control
function also covers the product \citep[pp.~428--429, 437]{wooldridge_control_2015}, and
\texttt{apexog} accepts this form as well. The interacted instruments are still needed in
it: without them the treatment's first stage, and with it the control function, is
misspecified whenever the instrument's effect on the treatment varies with the moderator.
The validated interaction configurations are
continuous-by-continuous and continuous-by-binary, with a continuous treatment. For a
binary endogenous treatment inside an interaction the control-function engine of
\texttt{apexog} computes the binary-by-binary contrast either with a linear first stage,
labelled a descriptive fitted contrast because a binary treatment cannot satisfy
Assumption~\ref{a:cf}, or with a probit first stage, which rests on the generalized-residual
approximation of Appendix~\ref{app:firststage}; in both, the product with the moderator is a
deterministic function of the two, so a separate first stage for the product does not define a
structural intervention. Neither configuration has been validated against a matching
data-generating process. The coherent version
models the treatment jointly, as the recursive bivariate probit does, with the
interaction in the outcome equation only; \texttt{apexog} offers it since version
0.22.0. There the AIE and its standard error equal those of \texttt{biprobit} followed
by \texttt{margins}, and in a bivariate-normal design the Wald interval at the true
violation covers the sample AIE at rates between $0.93$ and $0.97$ ($300$ replications
per cell). Its identification also rests on one error correlation for both values of the
moderator; under that restriction the interaction uses treatment variation the
instrument does not generate, and with an estimated correlation near zero the estimate is
close to that of a probit treating the treatment as exogenous (at a zero correlation the
likelihood factors into the two probits). Moderators with three or
more levels are not supported. On the definition of interaction effects in nonlinear models see
\citet{ai_interaction_2003} and \citet{greene_testing_2010}; on the analogous care
required in nonlinear difference-in-differences, \citet{puhani_treatment_2012}.

\subsection{The estimand: conditioning on the control function}
\label{app:estimand}

\begin{oaremark}[Single- versus double-average]\label{rem:single-double}
The expectation in \eqref{eq:ape-cts}--\eqref{eq:ape-bin} is over the joint sample
distribution of $(x_i,d_i,z_i,v_i)$, i.e.\ it conditions on the control function. This
is the estimand of \texttt{margins} after \texttt{cfprobit} and the one \texttt{apexog}
targets. It differs from the average structural function (ASF) derivative of
\citet{blundell_endogeneity_2003, wooldridge_unobserved_2005}, which integrates the
unobservable out and equals \eqref{eq:ape-cts} only when $\theta=0$. The distinction
matters for what a reported standard error refers to and is the reason our Monte Carlo
truth in Section~\ref{sec:mc} is the conditional-on-CF object.
\end{oaremark}

\section{Additional simulation detail}
\label{app:sims}

\subsection{The violation-to-estimate map, measured}
\label{app:map}

The map $m(\gamma_0;\gamma)$ of \eqref{eq:map} was measured directly in the design of
Section~\ref{sec:mc-design}: on a population proxy of $n=250{,}000$ with common
shocks across truths, the two-step estimator was refit on a $19\times9$ grid of
(assumed $\gamma_0$, true $\gamma$) pairs spanning
$[-0.9,0.9]\times[-0.6,0.6]$. Findings ((i) confirmed and (iv) measured by an independent
$R=300$ small-sample experiment at $n=2{,}000$ and $\gamma=0$): (i) the refit profile
$\gamma_0\mapsto m(\gamma_0;\gamma)$ is linear to within simulation precision (slope
$-0.2326$ from the grid, $-0.2333$ from the independent replications; measured
curvature $<10^{-4}$ across the full range); (ii) the offset sensitivity exceeds the
estimand sensitivity tenfold ($|\partial_1m|=0.233$ versus the grid-measured
$|\theta'(0)|=0.0236$, which matches Table~\ref{tab:collapse}'s single-run
$-0.023$ within simulation error); (iii) with the offset entered in structural units
the map's diagonal drifts from the truth with measured slope $+0.060$, consistent
within simulation precision with the implied
$|\partial_1m|(1/\tilde\sigma - 1)=0.058$, and entering the offset as
$(\gamma/\tilde\sigma)z$ restores
$m(\gamma;\gamma)=\theta(\gamma)$ to within $6\times10^{-4}$ at every tested truth,
the direct fingerprint of the units factor; (iv) the reported Murphy--Topel
standard error is well calibrated against the target-relative noise
($\mathrm{sd}$ of the estimator across replications $=0.94$ of the mean reported SE;
correlation of estimator noise with the sample-target's own noise $\approx0.05$),
which makes standard-error miscalibration an unlikely primary coverage mechanism in
this design. A reduced-model
simulator built from the measured map and noise components reproduces the observed
coverage profile and its plateau, and its ablations (linearized map, extra draws,
decorrelated noise, population target) move coverage by at most two points: in this
design the deficit travels through the units factor. Whether the map remains linear
and monotone under other links, discrete-change targets, or extreme baseline
probabilities is an empirical question. Two extensions of the check are informative.
Across $\rho\in\{0.3,0.6,0.8\}$ the continuous-$d$ refit map remains strictly
monotone and linear to within simulation precision (measured curvature $<10^{-4}$
throughout), with the offset slope $|\partial_1m|$ moving from $0.298$ at $\rho=0.3$
to $0.169$ at $\rho=0.8$: the linearity behind Proposition~\ref{prop:plateau}(b) is
not an artifact of one endogeneity strength, which is consistent with the plateau's
clean tracking of $\tilde\sigma$ in Table~\ref{tab:rhotrack}. The binary-$d$
discrete-change target behaves differently. In a binary-$d$ design (refits at nine offsets
on $\pm0.6$, $n=200{,}000$) its map is far steeper
(slope $\approx-1.31$) and carries genuine curvature ($\partial_1^2 m\approx0.58$),
so for such a target the linear-map closed form is only a first-order approximation
and coverage under a wide prior need not equal $\kappa_\infty$. Characterizing
coverage for curved maps, and thus for discrete-change and other strongly nonlinear
estimands, is the natural next step and the paper's principal remaining open
problem.

\begin{table}[t]
\centering
\caption{Agreement checks against independent benchmarks}
\label{tab:agree}
\footnotesize
\begin{tabular}{lll}
\toprule
Check & Result & Benchmark \\
\midrule
APE point and first-term SE at $\gamma=0$ & agree to $1.6\times10^{-9}$ & \texttt{cfprobit}+\texttt{margins} (GMM) \\
Two-term SE, sample and population & $10^{-14}$; $10^{-12}$ at offsets & formula-free oracle \\
Direct gradient $J_\pi$ & agrees to $10^{-12}$ & finite differences \\
Coefficients, all first stages, $\gamma=0$ & agree to $\ge11$ decimals & built-in two-step fits \\
MT/naive APE SE ratio at $\gamma=0.3$ & 1.162 vs.\ 1.146 & hand-coded GMM stack (coef.) \\
AIE point (manual decomposition) & agree to $1.2\times10^{-11}$ & \texttt{ginteff} oracle \\
AIE first-term SE & $0.98\times$ & 500-rep.\ full-pipeline bootstrap \\
Degenerate-prior band at $\gamma^*$ & $\le10^{-4}$ & oracle refit at $\gamma^*$ \\
Convolution interval, linear model & width ratio 0.995 & analytic $V+A\Omega A\trans$ \\
Cluster/robust APE SE ratio & 1.616 vs.\ 1.629 & native probit (coef.) \\
\bottomrule
\end{tabular}
\floatnote{Two-step estimator and its variance; MT = Murphy--Topel (first term of \eqref{eq:if-sample}). Details in Appendix~\ref{app:agree}.}
\end{table}

\subsection{Agreement checks and further robustness}
\label{app:agree}

\emph{Agreement checks (deterministic).} At $\gamma=0$, the two-step estimator equals
the built-in \texttt{cfprobit} point estimates exactly (coefficient differences
$\le10^{-11}$ across first-stage types), and the APE standard error built from the
first term of \eqref{eq:if-sample} equals the built-in \texttt{margins} standard error to
$1.6\times10^{-9}$, so \texttt{margins} omits the direct term. The full variance was checked
against a construction that uses no influence-function formula: with one instrument and a
linear first stage the second-stage regressors span the same columns as $(x_i,d_i,z_i)$, so the
fitted index does not depend on $\hat\pi$ and the APE equals
$\hat m\,(\hat b_d+\hat b_z/\hat\pi_z)$ from a probit of $y$ on $(x,d,z)$ and the first-stage
coefficient. One delta method on those stacked scores gives the exact first-order variance. On
$N=20{,}000$ simulated observations it matches \eqref{eq:full-effect-var} and the population variance to
$10^{-14}$ at $\gamma=0$ and to $10^{-12}$ at the true offset and at a misspecified one. The direct gradient matches finite
differences to $10^{-12}$ for linear and probit first stages, and a separate test differentiates
the stacked estimator numerically in every parameter for two endogenous regressors with
\texttt{cfinteract}, the continuous and the binary AIE, and clustering (agreement to $10^{-5}$).
The ratio of the two-term to the first-term standard error is $1.012$ in the baseline design at
$\gamma=0$, $1.014$ at the correct offset $0.3$, $1.004$ at a misspecified offset, $0.999$ with a
binary treatment and a probit first stage, $1.033$ at $\rho=0.9$, $1.001$ and $1.004$ in the
Catholic-schooling application, $1.000$ in the fertility application and $0.984$ in the clustered
clan application. At $\gamma=0.3$ the Murphy--Topel/naive ratio is $1.162$ for the APE
standard error, against $1.146$ for the standard error of the coefficient on $d$ in a hand-coded
GMM stack with the offset in the moment conditions.
The AIE point estimate equals the \texttt{ginteff} value on the manually decomposed
fit to $1.2\times10^{-11}$ in a real-data specification, and hand-coded oracles reproduce it to
$10^{-17}$ in the two-endogenous configuration; the first-term AIE SE is $0.978$ of a
500-replication full-pipeline bootstrap. Degenerate-prior checks: setting a point-mass prior at $\gamma^*$
reproduces the oracle refit at $\gamma^*$ to $10^{-4}$ or better, including asymmetric
multi-instrument $\mu$ vectors.

\emph{Designs of the interaction and two-instrument cells of Table~\ref{tab:ltz}.}
The AIE rows use an endogenous interaction of a continuous treatment with a binary
exogenous moderator: $x_i,z_i,v_i\sim N(0,1)$ and $x_{mi}\sim\text{Bernoulli}(0.5)$, mutually
independent, $d_i=0.7z_i+0.3x_i+0.2x_{mi}+v_i$ and
$y_i=\indic\{-0.3+0.5d_i+0.4x_i+0.3x_{mi}+0.4\,d_ix_{mi}+\gamma z_i+\varepsilon_i>0\}$ with
$\varepsilon_i=0.5v_i+e_i$, $e_i\sim N(0,0.75)$, so that $\tilde\sigma=\sqrt{0.75}$; $n=2{,}000$.
The endogenous regressors $d_i$ and $d_ix_{mi}$ are instrumented by $z_i$ and $z_ix_{mi}$ with
linear first stages, and the product of the two control functions enters the index. The
violation $\gamma\sim N(0,\sigma^2)$ sits on $z_i$ alone; the stated prior variance is
$\sigma^2$ on $z_i$ and $10^{-10}$ on $z_ix_{mi}$. The target is the continuous-by-binary AIE,
and the truth in each replication is its conditional-on-CF value at the true parameters
and the realized violation,
$n^{-1}\sum_i\{0.9\,\phinorm(\eta_i^{(1)})-0.5\,\phinorm(\eta_i^{(0)})\}/\tilde\sigma$, with
$\eta_i^{(r)}$ the index of \eqref{eq:cfprobit} at $x_{mi}=r$. The two-instrument rows use
the baseline design with $d_i=0.6z_{1i}+0.5z_{2i}+0.4x_i+v_i$ and independent violations on
both instruments, $\gamma_k\sim N(0,0.05^2)$ for the LTZ row and uniform on $[-0.2,0.2]$
in structural units for the UCI row, whose box of $\pm0.2$ is entered on the index scale.

\emph{The convolution matters.} In a linear-model oracle where the analytic
$V+A\Omega A\trans$ interval of \citet{conley_plausibly_2012} is available, the
draw-and-convolve interval matches it to a width ratio of $0.995$; omitting the
convolution (percentiles of the bias-adjusted points alone, i.e.\ a pure sensitivity
band) yields $0.703\approx1/\sqrt2$ of the correct width when prior-induced and
sampling variability are comparable, an understatement of uncertainty that motivates
reporting both objects separately, as \texttt{apexog} does.

\emph{Invariances of the plateau (as \eqref{eq:plateau} requires).} With $150$
prior draws per fit and the same $400$ replications and violations, the three LTZ-APE
scenarios of Table~\ref{tab:ltz} give $0.880$, $0.853$ and $0.330$ at $n=5{,}000$
against $0.900$, $0.855$ and $0.375$ at $n=2{,}000$. The plateau cell does not move, as a
plateau containing no $n$ requires; the transition and misstated cells fall by $2.0$ and
$4.5$ points (paired standard errors $1.8$ and $1.6$) because $\varrho$ grows with $\sqrt n$
at a fixed prior width, and \eqref{eq:kappaform} predicts falls of $1.3$ and $4.4$ points.
On the same $400$ replications at $n=2{,}000$, intervals built from
$150$ draws are about $2\%$ narrower than those built from $1{,}000$ (mean widths
$0.0582$ against $0.0593$ and $0.1834$ against $0.1865$ in the two $c=0.8$ cells),
the direction the order statistics imply. Width is the wrong diagnostic for the
draw count in any case: in the Gaussian linear limit experiment the coverage of a percentile
interval built from $B$ draws is
$(k_2-k_1)/(B+1)$ by the exchangeability of the target with the draws, $0.940$ at
$B=150$ and $0.949$ at $B=1{,}000$ (Section~\ref{sec:mc-design}), a property a
width comparison cannot see. In coverage the draw count shows as predicted: on the
same $400$ replications, data and seeds, moving from $150$ to $1{,}000$ draws raises
coverage by $1.3$ and $1.0$ points in the two $c=0.8$ cells of Table~\ref{tab:ltz} ($0.900$ to
$0.913$ and $0.855$ to $0.865$; paired standard errors $0.010$ and $0.010$), as a
simulation of the reduced model of Proposition~\ref{prop:plateau}(b) predicts
($+1.1$ points), and does not raise the misstated cell ($-1.3$ points, paired
standard error $0.7$). An
exogenous-moderator AIE Monte Carlo ($R=100$) gives $0.93$--$0.96$ across correctly
and incorrectly scaled priors. When the violation rides on an instrument orthogonal
to both moderators, the AIE is insensitive to $\gamma$ and even a misstated prior is
harmless; sensitivity flows only through channels correlated with the target.

\emph{Other checks.} Cluster-robust variances (scores summed within clusters in every
meat matrix) reproduce the native probit cluster/robust SE ratio to within $1\%$
($1.616$ for the APE against $1.629$ for the coefficient). Their coverage was checked in a design with a valid
instrument and strong cluster effects in the exogenous regressor, the instrument
and the latent index ($100$ clusters of $20$, $R=300$): the cluster-robust Wald
interval for the conditional-on-CF APE covers $0.943$ (Monte Carlo standard error
$0.013$), while the heteroskedasticity-robust interval, which ignores the
within-cluster dependence, covers $0.807$; mean interval widths are $0.047$ and
$0.033$. Non-Gaussian priors (uniform, $t$, $\chi^2$, gamma, Poisson,
and user-supplied empirical draws) run through the same convolution path; the
Gaussian prior specified as a distribution reproduces the closed-form
Gaussian path draw-for-draw under a common seed, and the uniform prior's coverage
under scale miscalibration is simulated in Section~\ref{sec:mc-ltz}. Multi-instrument
UCI unions are
conservative: at two instruments the union covers in $0.996$ of replications, although
the violations, drawn uniformly on $[-0.2,0.2]$ in structural units, put the true index
offset outside the $\pm0.2$ box in $355$ of the $1{,}000$ replications (three of the four
misses are among them).

\subsection{UCI coverage across first stages}
\label{app:uci}

In the experiments of Table~\ref{tab:uci} the true violation enters the structural
latent index while the box and the offsets are on the probit-index scale, so the
table's Wald column uses the index-scale offset $\gamma/\tilde\sigma$ for the linear
design; skipping that conversion costs Wald coverage ($0.536$ at $\gamma=0.2$ and
$0.014$ at $\gamma=0.5$, against $0.952$ and $0.939$ with it), the units gap of
Section~\ref{sec:coverage} seen directly in the union method.

Table~\ref{tab:uci} reports, for assumed support $\calG=[-0.3,0.3]$, three true
violations ($\gamma=0$, a valid instrument; $\gamma=0.2$, interior of $\calG$;
$\gamma=0.5$, well outside $\calG$) and five designs (the linear first stage, binary
$d$ under two data-generating processes, and the fractional-probit and Poisson first
stages of Appendix~B), four quantities per cell: the UCI coverage, the coverage of
the Wald interval at the true offset, the bias of the baseline ($\gamma_0=0$) point
estimate, and the mean width of the union. The UCI pattern is the theoretical one in
all fifteen cells: coverage $1.000$ whenever $\gamma\in\calG$ and $0.000$ when the
stated support is wrong, in $15{,}000$ replications. That pattern says nothing about
the standard errors, because the unions are $0.14$ to $1.1$ wide; the information is
in the other three columns. The Wald interval at the true offset tests the sizing of
the Murphy--Topel standard error at a nonzero violation directly. In the two designs
where the fit is correctly specified it is nominal ($0.946$, $0.952$ and $0.939$ for
the linear first stage; $0.951$, $0.954$ and $0.944$ for binary $d$ under the
CF-sufficient design; Monte Carlo standard error $0.007$--$0.008$), and the
truth-free ratio of the across-replication standard deviation of the estimate to the
mean reported standard error lies between $0.96$ and $1.03$ in every cell. The
lowest of these Wald coverages, the linear cell at $\gamma=0.5$ ($0.939$, ratio $1.03$), is
about $1.5$ Monte Carlo standard errors below nominal. The bias column settles the estimand question of
Remark~\ref{rem:single-double} numerically: the linear two-step is unbiased for the
conditional-on-CF APE ($0.0001$ at $\gamma=0$, within $0.0002$ at the true offsets)
and biased by $-0.0215$, or $13\%$, for the ASF derivative ($0.141$ against
$0.162$), which its Wald interval covers $4\%$ of the time. The reported standard error is
for the sample functional, and at the true offset it is right for either target. Away from the
true offset the two targets separate. At the $\gamma_0=0$ fit, when the true violation is $0.2$
or $0.5$, the across-replication standard deviation exceeds the mean reported standard error by
$5\%$ and $22\%$ in the linear design. An across-replication standard deviation measures
variation around a fixed number, the population pseudo-true value of that fit, and against the
population-target standard error of \eqref{eq:if-population}, computed in the same
replications, the ratios are $1.007$ and $1.005$ ($1.003$ and $0.998$ in $10{,}000$ further
replications); an interval built from it contains the Monte
Carlo mean of the estimates in $0.958$ and $0.948$ of the replications, against $0.944$ and
$0.902$ for the reported one. The gap is therefore the difference between $V_P$ and $V_S$, not
a defect of the sandwich. For binary $d$ the two standard errors agree within $0.4\%$ at every
fit (sampling standard deviation over either standard error $0.96$--$1.03$). The fractional-probit and Poisson designs show the same separation
on a smaller scale, which the $1{,}000$ replications of Table~\ref{tab:uci} are too few to
resolve: a ratio then has a Monte Carlo standard error of $0.02$, and at the $\gamma_0=0$ fit
with $\gamma=0.5$ those replications give $1.03$ (fractional probit) and $0.94$ (Poisson) against
the population-target standard error. In $10{,}000$ further replications of each of these
cells (Monte Carlo standard error $0.007$) the ratio in the fractional-probit design at
$\gamma=0.5$ is $1.06$
against the reported and $1.01$ against the population-target standard error. In the Poisson
design it is $1.01$ and $1.07$ at $\gamma=0.2$ and $0.5$ against the reported standard error and
$1.00$ and $1.01$ against the population-target one. The same replications test the reported
standard error against its own target. The standard deviation of the difference between the
estimate and the sample functional evaluated at the pseudo-true parameters of the $\gamma_0=0$
fit is $1.01$ times the mean reported standard error in both designs at $\gamma=0.5$, and the
reported interval contains that functional in $0.948$ (fractional probit) and $0.946$ (Poisson)
of the replications. The direct variance term carries weight in the Poisson cell: it raises the
reported standard error by $9\%$, and the first term alone falls $9\%$ short of that standard
deviation. An independent implementation of the linear, fractional-probit and Poisson designs
gives $1.00$ against the population-target standard error at $\gamma=0.5$ in $40{,}000$
replications of each. None of this touches the
UCI guarantee, which needs only the interval
at the true violation; it bears on the per-$\gamma$ bands and on LTZ intervals under wide
priors, where the sample-target standard error is applied at offsets far from the truth.

\begin{table}[t]
\centering
\caption{UCI coverage, Wald coverage at the true offset, bias and width}
\label{tab:uci}
\small
\begin{tabular}{@{}llcccc@{}}
\toprule
First stage (truth) & $\gamma$ & UCI & Wald & Bias & Width \\
\midrule
Linear (conditional-on-CF APE) & 0 & 1.000 & 0.946 & 0.000 & 0.165 \\
 & 0.2 & 1.000 & 0.952 & 0.056 & 0.157 \\
 & 0.5 & 0.000 & 0.939 & 0.131 & 0.145 \\
\addlinespace
Binary $d$, bivariate-normal design & 0 & 1.000 & 0.881 & $-0.024$ & 0.778 \\
\quad (structural discrete change) & 0.2 & 1.000 & n/a & 0.269 & 0.733 \\
 & 0.5 & 0.000 & n/a & 0.554 & 0.402 \\
\addlinespace
Binary $d$, CF-sufficient design & 0 & 1.000 & 0.951 & 0.001 & 0.787 \\
\quad (conditional-on-CF discrete change) & 0.2 & 1.000 & 0.954 & 0.280 & 0.743 \\
 & 0.5 & 0.000 & 0.944 & 0.559 & 0.433 \\
\addlinespace
Fractional probit (conditional-on-$u$ APE) & 0 & 1.000 & 0.897 & $-0.022$ & 1.138 \\
 & 0.2 & 1.000 & n/a & 0.317 & 1.081 \\
 & 0.5 & 0.000 & n/a & 0.743 & 0.964 \\
\addlinespace
Poisson (structural slope APE) & 0 & 1.000 & 0.953 & 0.000 & 0.334 \\
 & 0.2 & 1.000 & n/a & 0.097 & 0.332 \\
 & 0.5 & 0.000 & n/a & 0.232 & 0.329 \\
\bottomrule
\end{tabular}
\floatnote{Nominal level 0.95; $\calG=[-0.3,0.3]$, five grid points, $R=1{,}000$ per cell, $n=5{,}000$. The true violation enters the structural latent index; the box and the offsets are on the probit-index scale ($\gamma=0.2$ and $0.5$ are $0.25$ and $0.625$ index units in the linear design). The truth is named in the first column (Appendix~\ref{app:firststage}). Wald: coverage of the interval at the true offset ($\gamma/\tilde\sigma$ in the linear row); n/a where no offset makes the generalized-residual fit correctly specified. Bias: mean error of the $\gamma_0=0$ estimate. Width: mean length of the union. Monte Carlo SE of a Wald entry $\le0.010$; every replication inside the box covered and none outside.}
\end{table}

\subsection{The collapse experiment: invariance to instrument strength}
\label{sec:collapse}

The plateau formula \eqref{eq:plateau} makes a strong falsifiable prediction: at wide
priors, coverage depends on the design through the calibration ratio $c$ \emph{alone}.
Instrument strength moves the estimand sensitivity, the offset sensitivity, and the
sampling variance, but not $c=\tilde\sigma$, a function of the endogeneity
correlation only. A factorial that varies instrument strength at fixed prior width
should therefore find \emph{nothing}.

\begin{figure}[t]
\centering
\includegraphics[width=\textwidth]{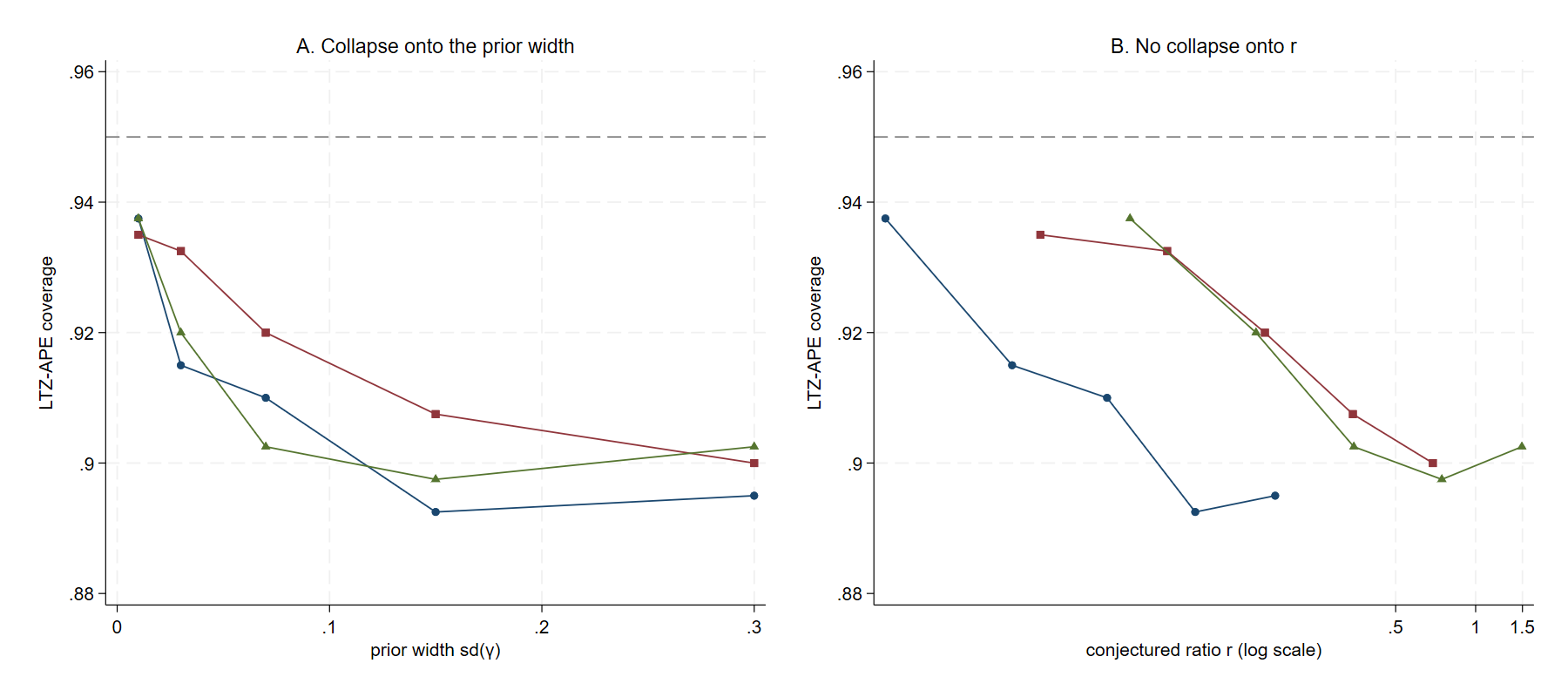}
\caption{The collapse experiment}
\label{fig:collapse}
\floatnote{Panel A: coverage against the prior width $\sigma$ (circles: $\pi_z=0.4$; squares: $\pi_z=0.8$; triangles: $\pi_z=1.2$). Panel B: the same fifteen cells against $r=|\theta'(0)|\sigma/\mathrm{SE}$ (log scale). Dashed line: nominal $0.95$. $R=400$ per cell, $n=2{,}000$, $c=\tilde\sigma=0.8$ throughout.}
\end{figure}

We ran that factorial, originally to test a signal-to-noise conjecture,
$r=|\theta'(0)|\,\sigma/\sqrt{V_\theta/n}$, built on the estimand sensitivity.
That conjecture used the wrong sensitivity. The
quantity that scales the prior's effect on the interval is the \emph{offset}
sensitivity $|\partial_1m|=0.233$, ten times $|\theta'(0)|=0.024$
(Section~\ref{sec:coverage}), so the tabulated $r$ understates the effective
signal ratio $\varrho$ of \eqref{eq:kappaform} by an order of magnitude, and the cells
reach deep into the plateau regime rather than probing its onset. What the experiment
delivers is a check of invariance to instrument strength. Instrument strength
$\pi_z\in\{0.4,0.8,1.2\}$ moves $\theta'(0)$ across $\{-0.013,-0.023,-0.030\}$ and
the APE standard error across $\{0.021,0.010,0.006\}$, crossed with structural-units
prior widths $\sigma\in\{0.01,0.03,0.07,0.15,0.30\}$ ($c=0.8$ throughout), at
$R=400$ per cell. Table~\ref{tab:collapse} reports all fifteen coverages;
Figure~\ref{fig:collapse} plots them against $\sigma$ (panel~A) and against the
conjectured $r$ (panel~B).

\begin{table}[t]
\centering
\caption{The collapse experiment: coverage by instrument strength and prior width}
\label{tab:collapse}
\small
\begin{tabular}{lcc ccccc}
\toprule
& & & \multicolumn{5}{c}{Prior width $\sigma$} \\
\cmidrule(lr){4-8}
Instrument & $\theta'(0)$ & SE & 0.01 & 0.03 & 0.07 & 0.15 & 0.30 \\
\midrule
Weak ($\pi_z=0.4$)   & $-0.013$ & 0.021 & 0.938 & 0.915 & 0.910 & 0.893 & 0.895 \\
Medium ($\pi_z=0.8$) & $-0.023$ & 0.010 & 0.935 & 0.933 & 0.920 & 0.908 & 0.900 \\
Strong ($\pi_z=1.2$) & $-0.030$ & 0.006 & 0.938 & 0.920 & 0.903 & 0.898 & 0.903 \\
\addlinespace
$r$ range across rows & & & 8$\times$ & 8$\times$ & 8$\times$ & 8$\times$ & 8$\times$ \\
Within-column spread & & & 0.003 & 0.018 & 0.018 & 0.015 & 0.008 \\
\bottomrule
\end{tabular}
\floatnote{LTZ-APE coverage, nominal level 0.95; $R=400$ per cell (Monte Carlo SE $\approx0.015$), $n=2{,}000$, $150$ prior draws per fit, structural-units priors ($c=\tilde\sigma=0.8$ in every cell). $\theta'(0)$ and SE: design-level estimand sensitivity and standard error from a single baseline fit; $r=|\theta'(0)|\sigma/\mathrm{SE}$. Summary rows: max/min ratio of $r$ and max$-$min coverage within each column, from unrounded coverages.}
\end{table}

The prediction is not contradicted. Within each prior width, coverage
is flat across instrument strengths (mean within-column spread $0.0120$,
indistinguishable from binomial Monte Carlo noise $\approx0.015$); across prior
widths it moves by $0.04$, tracing the transition of \eqref{eq:kappaform} into the
common plateau. Regressing the fifteen coverages on the prior width and the
instrument strength gives a width gradient of $-0.112$ ($p=0.001$) and an
instrument-strength gradient of $0.003$ ($p=0.78$), statistically indistinguishable from
zero; the design's per-cell resolution of $\approx0.015$ bounds any undetected effect well
below the width gradient. Given $\sigma$, nothing the instrument does, whether to
the sensitivity, to the sampling variance, or to any signal-to-noise ratio, moves
coverage detectably. That is what a plateau depending only on $(\alpha, c)$, with
$c$ common to all cells, requires, but the factorial is weak evidence for it. By the
identity \eqref{eq:affine-profile} the offset sensitivity is $\hat m/|\hat\pi_z|$, and
the standard error scales roughly with $1/|\pi_z|$ as well, so the ratio
$\varrho/\sigma$ of \eqref{eq:kappaform} is about $22$ to $24$ at the three
instrument strengths. The closed form therefore predicts coverages within a quarter of a
point of one another across instrument strengths at every prior width, in the transition
as well as in the plateau. Flatness across the rows of Table~\ref{tab:collapse} confirms that
prediction without separating the plateau's freedom from the design from this
cancellation; the sample-size comparison of Appendix~\ref{app:agree}, which does move
$\varrho$, is the sharper check.

\subsection{What the linear-probability shortcut costs}
\label{sec:mc-lpm}

Finally we quantify the cost of the tempting alternative: treat the binary outcome as
a linear probability model and run the linear machinery of
\citet{conley_plausibly_2012} via \texttt{plausexog}. Motivated by the
skewed-regressor critique of \citet{chen_another_2023} for OLS, one might expect the
shortcut to fail when the endogenous regressor is skewed. The IV setting is kinder
more forgiving in one respect and less forgiving in another. With a linear first
stage, two-stage least squares
equals the reduced-form-to-first-stage ratio, and for a (near-)Gaussian
\emph{instrument} an application of Stein's lemma shows this ratio is a
probability-scale average-derivative effect, numerically within about one percent of the
APE in our Gaussian-instrument designs, regardless of the regressor's distribution (Appendix~D.4 gives
the short derivation). Under a linear first stage, skewness of $d$ alone therefore
cannot break the shortcut: in a design
with a heavily skewed endogenous regressor (skewness $2.1$) and a Gaussian instrument,
the LPM-IV estimate errs by $-1.1\%$ and the CF-probit by $-1.0\%$. (Instrument
distributions we have not simulated, notably binary instruments, remain an open
case.) What breaks the
shortcut is a skewed \emph{instrument}. Table~\ref{tab:lpm} reports a design
identical in every respect except that the instrument is a centered standard lognormal
(skewness $\approx6.2$): the true APE is $0.105$, the CF-probit delivers $0.104$
($-0.4\%$), and the LPM-IV delivers $0.084$, a $20\%$ bias, at $n=200{,}000$, so
the gap is bias, not noise.

\begin{table}[t]
\centering
\caption{The LPM shortcut under a skewed instrument}
\label{tab:lpm}
\footnotesize
\begin{tabular}{lcc cc}
\toprule
& \multicolumn{2}{c}{CF-probit (\texttt{apexog})} & \multicolumn{2}{c}{LPM (\texttt{plausexog})} \\
\cmidrule(lr){2-3}\cmidrule(lr){4-5}
& Bounds & Covers truth? & Bounds & Covers truth? \\
\midrule
Point estimate ($\gamma=0$, bias) & $0.1044$ $(-0.4\%)$ &  & $0.0836$ $(-20.3\%)$ &  \\
$b=0.005$ & $[0.097,\,0.112]$ & yes & $[0.076,\,0.091]$ & \textbf{no} \\
$b=0.010$ & $[0.091,\,0.118]$ & yes & $[0.070,\,0.097]$ & \textbf{no} \\
$b=0.020$ & $[0.078,\,0.131]$ & yes & $[0.058,\,0.110]$ & yes \\
$b=0.030$ & $[0.066,\,0.143]$ & yes & $[0.045,\,0.122]$ & yes \\
\midrule
Control: Gaussian instrument & \multicolumn{2}{c}{$0.1141$ $(+1.3\%)$} & \multicolumn{2}{c}{$0.1139$ $(+1.1\%)$} \\
\bottomrule
\end{tabular}
\floatnote{Centered lognormal instrument (skewness $\approx6.2$), valid ($\gamma=0$); one replication at $n=200{,}000$, so the discrepancies are bias. True APE $=0.1048$; the Gaussian-instrument control row is a separate design (true APE $0.1127$; point estimates only). Bounds: UCI over $|\gamma|\le b$ on the probability scale (\texttt{gunits(pr)}, nine grid points). First row: point estimates with percent bias; bold: bounds that exclude the truth.}
\end{table}

The decisive rows are the tight-prior ones. A sensitivity analysis is most valuable
precisely when the researcher believes the violation is small; but for
$b\le0.01$, where the instrument is believed nearly valid, the LPM bounds, centered
on a $20\%$-biased point, \emph{exclude} the true APE entirely, while the CF-probit
bounds contain it at every prior width. (These containment verdicts isolate
deterministic bias in a single large-$n$ replication; at $n=200{,}000$, sampling
noise is negligible relative to the $20\%$ gap, though the $b=0.01$ row sits nearer
the boundary.) The bias hides only under priors loose enough
to cover it by accident. Two caveats complete the picture. With a Gaussian
instrument both methods are fine (bottom row): the shortcut's failure is about the
instrument's distribution, not about nonlinearity per se. Meanwhile, the one configuration
in which regressor skewness does matter, a genuinely nonlinear first stage with
$d=\exp(0.5z+0.5u)$, biases the \emph{linear-control-function} probit estimator worse
($+35.5\%$) than the LPM ($-9.2\%$): that is a first-stage misspecification problem,
common to both approaches, not an argument between them.

\section{Computation}
\label{app:comp}

\subsection{Gradients and the influence function}
\label{app:jac}

Let $r_i$ collect the second-stage regressors (elements of $g_i$), index the
entries of $\vartheta$ by $\ell$, and write
$\phinorm'(\eta)=-\eta\,\phinorm(\eta)$. For the continuous APE
$\hat\theta=n^{-1}\sum_i\phinorm(\hat\eta_i)\hat\delta$,
\[
\frac{\partial\hat\theta}{\partial\vartheta_\ell}
= n^{-1}\sum_i\Big[\phinorm'(\hat\eta_i)\,r_{\ell i}\,\hat\delta
+ \phinorm(\hat\eta_i)\,\indic\{\vartheta_\ell=\delta\}\Big].
\]
For the binary-$d$ APE \eqref{eq:ape-bin},
$\partial\hat\theta/\partial\vartheta_\ell
= n^{-1}\sum_i[\phinorm(\hat\eta^{(1)}_i)r^{(1)}_{\ell i}-\phinorm(\hat\eta^{(0)}_i)r^{(0)}_{\ell i}]$
with $r^{(d)}$ evaluated at $d_i=d$. For the continuous-by-continuous AIE
\eqref{eq:aie-ctscts}, with $a_i,b_i$ as defined there,
\[
\frac{\partial\widehat{\AIE}}{\partial\vartheta_\ell}
= n^{-1}\sum_i\Big\{\phinorm'(\hat\eta_i)r_{\ell i}\big(\hat\delta_{jk}-\hat\eta_i a_ib_i\big)
+ \phinorm(\hat\eta_i)\Big[\indic\{\vartheta_\ell=\delta_{jk}\} - r_{\ell i}a_ib_i
- \hat\eta_i\big(\tfrac{\partial a_i}{\partial\vartheta_\ell}b_i
+ a_i\tfrac{\partial b_i}{\partial\vartheta_\ell}\big)\Big]\Big\},
\]
where $\partial a_i/\partial\vartheta_\ell=\indic\{\vartheta_\ell=\delta_j\}+r_{ki}\indic\{\vartheta_\ell=\delta_{jk}\}$
and symmetrically for $b_i$: the generic chain-rule term, the direct derivative of
the leading $\delta_{jk}$, and corrections at the two own coefficients and the
interaction coefficient. For the continuous-by-binary AIE, with the moderator
$r_k\in\{0,1\}$ and $\eta_i^{(k)}$ denoting the index at $r_{ki}=k$,
\begin{align*}
\widehat{\AIE} &= n^{-1}\sum_i\big[\phinorm(\hat\eta^{(1)}_i)(\hat\delta_j+\hat\delta_{jk})
- \phinorm(\hat\eta^{(0)}_i)\,\hat\delta_j\big], \\
\frac{\partial\widehat{\AIE}}{\partial\vartheta_\ell}
&= n^{-1}\sum_i\Big[\phinorm'(\hat\eta^{(1)}_i)r^{(1)}_{\ell i}(\hat\delta_j+\hat\delta_{jk})
- \phinorm'(\hat\eta^{(0)}_i)r^{(0)}_{\ell i}\hat\delta_j \\
&\qquad\qquad
+ \phinorm(\hat\eta^{(1)}_i)\big(\indic\{\vartheta_\ell=\delta_j\}+\indic\{\vartheta_\ell=\delta_{jk}\}\big)
- \phinorm(\hat\eta^{(0)}_i)\indic\{\vartheta_\ell=\delta_j\}\Big].
\end{align*}
For the binary-by-binary AIE, with $\eta^{(j,k)}_i$ the index at
$(r_{ji},r_{ki})=(j,k)$,
\begin{align*}
\widehat{\AIE} &= n^{-1}\sum_i\big[\Phinorm(\hat\eta^{(1,1)}_i)-\Phinorm(\hat\eta^{(1,0)}_i)
-\Phinorm(\hat\eta^{(0,1)}_i)+\Phinorm(\hat\eta^{(0,0)}_i)\big], \\
\frac{\partial\widehat{\AIE}}{\partial\vartheta_\ell}
&= n^{-1}\sum_i\sum_{(j,k)} (-1)^{2-j-k}\,\phinorm(\hat\eta^{(j,k)}_i)\,r^{(j,k)}_{\ell i}.
\end{align*}
These are the gradients $J_\vartheta$ of \eqref{eq:jac}, evaluated at the fitted parameters
and sample covariates.

\emph{The direct first-stage gradient.} Each per-unit effect depends on $\pi$ only through the
index, so with $D_i=\partial q_i/\partial(\text{index shift})$ and control $c_i(\pi)$,
\[
J_\pi=\E\big[D_i\;(\partial\eta_i/\partial c_i)\;\partial c_i/\partial\pi\big],
\]
where $\partial\eta_i/\partial c_i=\tcf$, plus each product coefficient times the partner
control when products of control functions enter the index, and $\partial c_i/\partial\pi=-w_i$ for a linear first stage
(for a nonlinear first stage, the derivative of its generalized residual). With several first
stages the gradient is stacked over them. In the notation above, $D_i$ is
$\hat\delta\,\phinorm'(\eta_i)$ for the continuous APE;
$\phinorm(\eta_i^{(1)})-\phinorm(\eta_i^{(0)})$ for the binary APE;
$\phinorm'(\eta_i)(\delta_{jk}-\eta_ia_ib_i)-\phinorm(\eta_i)a_ib_i$ for the
continuous-by-continuous AIE;
$\phinorm'(\eta^{(1)}_i)(\delta_j+\delta_{jk})-\phinorm'(\eta^{(0)}_i)\delta_j$ for the
continuous-by-binary AIE; and $\sum_{(j,k)}(-1)^{2-j-k}\phinorm(\eta^{(j,k)}_i)$ for the
binary-by-binary AIE.

\emph{Implementation and clustering.} Compute $\hat\psi_{1i}$ and
$\hat\psi_{2i}=-\hat H^{-1}(\hat s_i+\hat G\hat\psi_{1i})$, and form
$\hat\ell_{Si}=\hat J_\vartheta\trans\hat\psi_{2i}+\hat J_\pi\trans\hat\psi_{1i}$ or
$\hat\ell_{Pi}=\hat q_i-\hat\theta+\hat\ell_{Si}$. Consistent variance estimates are their
empirical variances. For independent clusters indexed by $h$, a cluster sandwich for the
variance of $\hat\theta$ is $n^{-2}\sum_h(\sum_{i\in h}\hat\ell_i)^2$, with the usual optional
finite-cluster correction (\texttt{apexog}'s two-step engine applies none; with
\texttt{firststage(biprobit)} the standard errors are Stata's maximum-likelihood ones, which
carry $n/(n-1)$ or $G/(G-1)$). Its justification requires a
many-cluster limit, appropriate moment conditions and no dominating cluster; the i.i.d.\ theorem
alone does not establish clustered coverage. \texttt{apexog} reports $\hat V_S$.

\emph{Recursive bivariate probit.} This estimator uses its joint parameter influence function
$\psi_{Bi}$. If its structural probability contrast is $q_{Bi}(\xi)$, the sample-target
influence is $\E[\partial_\xi q_{Bi}]\trans\psi_{Bi}$; for the population average treatment
effect add $q_{Bi}-\E[q_{Bi}]$ inside that influence function. Agreement of point estimates
across estimators does not establish agreement of these target-specific standard errors.

\emph{Counterfactual offsets.} At an offset other than the true violation the information
identity fails, and the derivatives must be the full score derivatives. With
$r_i=\lambda(\eta_i)(y_i-\Phinorm(\eta_i))$ and
$h_i=\lambda'(\eta_i)(y_i-\Phinorm(\eta_i))-\lambda(\eta_i)\phinorm(\eta_i)$,
\[
\partial_\vartheta s_i=h_i g_i g_i\trans,\qquad
\partial_{\pi\trans}s_i=-\tcf h_i g_iw_i\trans-r_i e_{\mathrm{cf}}w_i\trans,
\]
whose sample averages estimate $H$ and $G$, with $e_{\mathrm{cf}}$ selecting the control
coefficient. The effect's direct first-stage derivative is still required.

\subsection{The sensitivity $A$}
\label{app:A}

The pseudo-true second-stage parameter at assumed offset $\gamma_0$ solves
$\E[s(\vartheta;\gamma_0)]=0$; the implicit function theorem gives
$d\vartheta^*/d\gamma_0 = -H^{-1}\E[\partial s_i/\partial\gamma_0]$ with
$\E[\partial s_i/\partial\gamma_0]=-\E[\omega(\eta_i)g_iz_i\trans]$ (the same algebra
as the Hessian, with $z_i$ replacing $g_i$ on one side). Adding the direct dependence
of the effect functional on the offset,
\[
\frac{\partial m}{\partial\gamma_0}
= \E[\phinorm'(\eta_i)z_i]\,\delta
- J_\vartheta\trans\,\E[\omega(\eta_i)g_ig_i\trans]^{-1}\E[\omega(\eta_i)g_iz_i\trans],
\qquad A = -\,\frac{\partial m}{\partial\gamma_0},
\]
for the continuous APE; other targets replace the direct term by the corresponding
$\partial q/\partial\gamma_0$. In general, with $K=\E[\partial s_i/\partial\gamma\trans]$ at
$(0,0)$, $A=J_\vartheta\trans H^{-1}K-\E[\partial q_i/\partial\gamma\trans]$. The first-stage fit
does not change with the assumed offset, so its estimation uncertainty enters $V_\theta$, not
an additional offset derivative in $A$. In practice $A$ need never be computed: the refit
$\hat\theta(\gamma_0)$ embodies the exact finite-sample map, and $A$ appears only in
the theory. Two magnitudes should not be conflated: $A=-\partial_1m$ (how hard the
\emph{estimate} responds to the assumed offset; $0.233$ in the design of
Section~\ref{sec:mc}) and $\theta'=\partial_1m+\partial_2m$ (how much the
\emph{estimand} moves with the actual violation; $-0.024$ there). The prior's effect
on the LTZ interval scales with the former; sensitivity intuitions built on the
latter understate it by an order of magnitude.

\subsection{Algorithms}
\label{app:alg}

\emph{Software.} The Stata and R code was developed with the AI assistance disclosed in the
acknowledgments of the main paper; Appendix~\ref{app:agree} reports the checks of the
estimator against \texttt{cfprobit}, \texttt{margins} and \texttt{ginteff} and against
constructions that use no influence-function formula.

\emph{UCI.} Grid $\calG$ with $n_g$ points per instrument ($n_g^{k_z}$ nodes); at each node
refit with the offset and record
$\hat\theta\pm z_{1-\alpha/2}\hat\sigma$; report the hull \eqref{eq:uci}.
The implementation refuses grids exceeding the platform's matrix limits with a
pointer to the LTZ method (which scales in draws, not $g^{k_z}$).
\emph{Grid accuracy.} In practice the extrema in \eqref{eq:uci} are computed on a finite grid over $\calG$. Under
R4 the maps $\gamma_0\mapsto(\hat\theta(\gamma_0),\hat\sigma(\gamma_0))$ are continuous,
so the grid approximation error vanishes as the mesh shrinks; with $k_z$ instruments the
grid has $\text{(points)}^{k_z}$ nodes, and \texttt{apexog} guards against infeasible
grids and offers the LTZ method, which draws rather than grids, for large $k_z$
(Appendix~D).

\emph{LTZ.} Draw $\gamma^{(j)}\sim F$, $j=1,\dots,B$ (default $B=1{,}000$; seeded for
reproducibility); refit at each draw for $(\hat\theta_j,\hat\sigma_j)$; convolve
$\theta^{*}_j=\hat\theta_j+\hat\sigma_j\epsilon_j$ with a full $B$-vector of
independent standard normals drawn regardless of any failed fits (so the random-number
state is reproducible); report the empirical $\alpha/2,1-\alpha/2$ quantiles of
$\{\theta^*_j\}$ as the interval and the quantiles of $\{\hat\theta_j\}$ as the
prior-only sensitivity band. Failed draws (rare; zero in $3{,}000$ stress draws) are
dropped from both. With $B$ draws and quantile positions
$k_1=\mathrm{round}(\alpha B/2)$ and $k_2=\mathrm{round}((1-\alpha/2)B)$, the
coverage of the percentile interval in the Gaussian linear limit experiment, where target and
draws are exchangeable, is $(k_2-k_1)/(B+1)$: $0.949$ at
the default $B=1{,}000$ and $0.940$ at $B=150$.

\emph{Units.} The offset lives in probit-index units (equivalently, ``index units''
or ``offset units''; we use the three terms interchangeably). Three elicitation
scales map onto it. (i) \emph{Probability scale}: with the baseline fit's
$\hat{\bar m}=n^{-1}\sum_i\phinorm(\hat\eta_i)$, a violation stated as a shift of
$\Prob(y=1)$ per unit of $z$ converts by $\gamma\mapsto\gamma/\hat{\bar m}$
($\Omega\mapsto\Omega/\hat{\bar m}^2$), a first-order conversion held fixed at the
baseline; the output reports $\hat{\bar m}$. This scale references neither latent
index; the conversion's own linearization error is second-order in the prior width.
(ii) \emph{Probit-index units}: used directly. (iii) \emph{Structural latent units}
(the index whose total error has unit variance): divide by
$\hat{\tilde\sigma}=(1+\hat\sigma_v^{2}\hat\tcf^{\,2})^{-1/2}$ from
\eqref{eq:sigtilde} (variances by $\hat{\tilde\sigma}^{2}$) before use; omitting the
conversion is the $c=\tilde\sigma$ miscalibration priced by
Proposition~\ref{prop:plateau}. The conversion is derived for the linear first
stage; with a generalized-residual first stage the simulations of Appendix~B put
the correct offset at the unscaled structural $\gamma$, and $\hat{\tilde\sigma}$,
which is also a baseline-fit quantity that moves with the violation, is not the
relevant factor. Because each refit subtracts the assumed violation,
the percentile interval of the convolved refits is the pivot interval of
Corollary~\ref{cor:exact} for any prior shape, symmetric or not (Appendix~A); the
asymmetric priors offered by \texttt{distribution()} (e.g., $\chi^2$ or gamma) need
no separate treatment. In a linear-model check with centered $\chi^2$, exponential
and gamma priors the percentile-of-refits interval covers at $0.950$, $0.950$ and
$0.949$ (replication files).

\subsection{Why the LPM-IV estimate needs a well-behaved instrument}
\label{app:stein}

Partial the exogenous controls out and let $z$ be scalar with mean zero. The IV
estimand is $\beta_{IV}=\Cov(y,z)/\Cov(d,z)$. With the linear first stage
$d=\pi z+v$, $z\independent v$, the denominator is $\pi\Var(z)$. For the numerator,
$\Cov(y,z)=\Cov(\E[y\given z],z)$, and under the model
$\E[y\given z]=\Phinorm\big((c_0+(\delta\pi+\gamma)z)/\varsigma\big)$ with
$\varsigma^2=\Var(\delta v+\varepsilon)$ and $c_0$ the partialled intercept. If $z$
is \emph{Gaussian}, Stein's lemma gives
$\Cov(h(z),z)=\Var(z)\,\E[h'(z)]$, so at $\gamma=0$
\[
\beta_{IV}
= \frac{\Var(z)\cdot\pi\delta\,\E\big[\phinorm\big((c_0+\delta\pi z)/\varsigma\big)\big]/\varsigma}{\pi\Var(z)}
= \delta\,\E\big[\phinorm(\cdot)\big]/\varsigma ,
\]
a $\phinorm$-weighted average derivative of the response probability. It equals the
population APE of Definition~\ref{def:ape}: writing $\E[y\given z]=\E_v[\Phinorm(\eta_i)]$ with
$\eta_i$ the conditional-on-control index and differentiating under the expectation gives
$\beta_{IV}=\E[(\delta/\tilde\sigma)\phinorm(\eta_i)]$, \emph{whatever} the
distribution of $d$ (the regressor's skewness never entered). In the Gaussian-instrument
designs of Appendix~\ref{sec:mc-lpm} the two estimates differ by less than $1\%$. The identity uses the
instrument's Gaussianity exactly once, in Stein's lemma; for a skewed instrument
$\Cov(h(z),z)\ne\Var(z)\E[h'(z)]$ in general, the ratio loses its average-derivative
interpretation, and the $20\%$ bias of Table~\ref{tab:lpm} results. This is why the
skewed-\emph{regressor} critique of \citet{chen_another_2023}, developed for OLS,
transmits to the IV setting through the instrument rather than the regressor.

\section{Additional details on the applications}
\label{app:apps}

\subsection{Exact and first-order probability scales}
\label{app:apps-scale}

Both tipping values of Table~\ref{tab:apps} are read off the union grid by linear
interpolation between adjacent grid points; because the refits are monotone in a
scalar violation, the first is the violation at which the union interval of
Proposition~\ref{prop:uci} with support $[0,\gamma]$ (or $[-\gamma,0]$) stops
excluding zero. The tipping values state the violation on the exact probability scale
of Section~\ref{sec:app}, which \texttt{apexog} reports for every refit since
version 0.20; the \texttt{gunits(pr)} option converts
at first order, dividing $\gamma$ by the mean normal density of the outcome index at
the baseline fit. With a continuous instrument, a linear first stage and as many
instruments as endogenous regressors the two coincide: the offset then lies in the
span of the regressors and control functions, so the refit leaves the fitted index
unchanged (the clan application is of this kind). With a binary instrument they
differ through the curvature of $\Phi$ and the movement of the refit; at the tipping
values the exact values are up to $3\%$ larger in the fertility data and $4\%$ to
$11\%$ smaller in the Catholic-schooling data, whose outcome is $93\%$ ones. On the
first-order scale the Catholic-schooling probit values would read $4\%$ to $13\%$
higher, the gap being largest for the bivariate probit.

\subsection{Catholic schooling}
\label{app:apps-catholic}

\citet{van_kippersluis_beyond_2018} combine the standard error of the
zero-first-stage reduced form with the remaining sample's by a
normalized-difference rule to set the prior variance, which gives
$p\sim N(0.036,\,0.0018^{2})$ on the probability scale. This prior was not entered
through \texttt{gunits(pr)}, whose first-order conversion would center the offset at
$0.036/0.130=0.277$. We solved for the index offset whose exact probability effect in the
refitted bivariate probit is $0.036$, which is $0.310$, and divided $0.0018$ by the slope
of the exact map at that point ($0.122$), which gives a standard deviation of $0.0148$.
The prior analyzed is therefore $\gamma_0\sim N(0.310,\,0.0148^{2})$ in index units, a
plug-in calibration that treats the estimated map as known. The exact probability effects
at its $1{,}000$ refits have mean $0.0361$, standard deviation $0.0018$ and
$2.5\%$ and $97.5\%$ quantiles $0.0326$ and $0.0396$ (the Gaussian values are
$0.0325$ and $0.0395$). The central $95\%$ of the point estimates under the prior alone lie
between $-0.076$
and $-0.027$, and the LTZ interval is $[-0.30,\,0.20]$; the LPM point estimate becomes
$-0.052$ ($0.040$). Both are negative and far from significant.

\emph{The refit path.} The standard error of the bivariate-probit APE grows from $0.007$
at $\gamma_0=0$ to $0.124$ at the benchmark offset, which is why the LTZ interval is so
much wider than the prior-only band. The refits at the nine offsets examined, from $0$
to $0.75$, converge in two to five iterations, and the estimated error correlation stays
in the interior: $\hat\rho$ moves
from $-0.40$ to $0.35$ at the benchmark offset, with a $95\%$ interval of
$[-0.28,\,0.77]$ there. Two things grow. The standard error of the treatment coefficient
rises from $0.27$ to $0.68$, and the delta-method factor that multiplies it, the mean
density of the outcome index with the treatment switched on, rises from $0.015$ to
$0.18$: at $\gamma_0=0$ the treated index sits against the ceiling of an outcome that is
$93\%$ ones, and the refit moves it off that ceiling. Their product at the benchmark,
$0.124$, is close to the standard error of the APE there. Most of the widening is thus the
nonlinearity of the probability scale rather than a failing fit, but the treatment
coefficient and the error correlation are less precisely determined at large offsets, and
the normal approximation behind the Wald intervals there has not been checked by
simulation.

The calibration is borrowed rather than re-estimated, and that limits the
comparison with their result. Their sample of $8{,}802$ students carries controls
the public extract lacks (household composition, parents' marital status,
urbanicity), and the extract does not identify the public-school eighth graders on
whom the direct effect is measured. The difference shows in the reduced form:
theirs is $0.037$, almost exactly the measured direct effect, which is why their
estimate lands near zero; ours is $0.027$, so subtracting the same $0.036$
overshoots and turns the sign. The baselines differ as well, for a reason that
does not rank the estimators: with a binary treatment and a binary instrument,
2SLS estimates, under the IV and monotonicity assumptions, the effect for students whose
Catholic schooling the instrument moves, whereas the bivariate probit and the
probit-first-stage control function estimate a population discrete change under their
parametric assumptions (Section~\ref{sec:app-ae}); the linear-first-stage row is a
descriptive contrast. The 2SLS coefficient is twice the probit APEs here.

\subsection{Angrist--Evans}
\label{app:apps-ae}

The 1980 Census extract was rebuilt from the authors' public replication files,
following their SAS program line by line. Under heterogeneous effects the population
average treatment effect of the bivariate probit and the complier effect of 2SLS
differ. The control-function probit with a linear first stage, whose first-stage
residual is the LPM residual, agrees with 2SLS numerically in this sample. The agreement
does not establish that it estimates a complier effect: a binary treatment cannot satisfy
Assumption~\ref{a:cf}, and the row is a descriptive fitted contrast
(Section~\ref{sec:app}).

On the plausibility of a negative direct effect: the channel
\citet{rosenzweig_natural_2000} document runs through child costs. In rural Indian
data, clothing for a child with an older sibling of the same sex costs about a
fifth less, and a cheaper child could lower the mother's work through an income
effect, the threatening direction; they do not estimate the labor-supply effect of this
channel.
The Dutch direct effect of two boys on employment quoted in Section~\ref{sec:app-ae}
is from Table~B.1 of \citet{van_kippersluis_beyond_2018}.

The LTZ interval under a symmetric $N(0,\,0.0025^{2})$ prior (a quarter of a point),
entered through \texttt{gunits(pr)} and therefore a mean-zero Gaussian prior with standard
deviation $0.0025/0.373=0.0067$ on the index scale ($300$ draws),
is $[-0.28,\,-0.14]$ for the bivariate-probit APE, against a prior-only band of
$[-0.27,\,-0.16]$: at this width the prior contributes more than sampling noise (the
$\gamma=0$ Wald interval is $0.080$ wide, the prior-only band $0.109$). With both
instruments of the original paper (same-sex and twin births) and a box of $\pm0.01$
on each, the LPM union interval is $[-0.164,\,-0.015]$ and the bivariate-probit union
interval $[-0.190,\,-0.062]$; both exclude zero.

\subsection{Confucian clans}
\label{app:apps-clan}

The specification is that of \citet{chen_banking_2022}, Table~9, column~2. Individual
controls are sex, age, minority status, rural registration, education and number of
children; prefectural controls are distances to the coast, to navigable rivers and to
treaty ports, provincial capital, GDP per capita, Protestant missionaries in 1920,
modern banks in 1936, and the density of imperial degree holders; standard errors are
clustered by prefecture. With an outcome mean of $14\%$ and a continuous treatment the
LPM and the probit agree closely, and with a continuous instrument, a linear first
stage and exact identification the exact and first-order probability scales coincide.

\end{document}